\documentclass[%
 aip,
 amsmath,amssymb,
 reprint,%
fleqn,
floatfix
]{revtex4-1}

\DeclareRobustCommand{\threedots}{%
  \mathrel{\vbox{\baselineskip.65ex\lineskiplimit0pt\hbox{.}\hbox{.}\hbox{.}}}%
}

\usepackage{graphicx}
\usepackage{dcolumn}
\usepackage{bm}
\usepackage[T1]{fontenc}
\usepackage{mathptmx}
\usepackage{etoolbox}
\usepackage{annotate-equations}
\usepackage{amsmath,nccmath}
\usepackage{hyperref}

\makeatletter
\def\@email#1#2{%
 \endgroup
 \patchcmd{\titleblock@produce}
  {\frontmatter@RRAPformat}
  {\frontmatter@RRAPformat{\produce@RRAP{*#1\href{mailto:#2}{#2}}}\frontmatter@RRAPformat}
  {}{}
}%
\makeatother

\newcommand{\lgradb}{$L_{ \nabla \mathbf{B}}$}
\newcommand{\minlgradb}{$\mathrm{min}(L_{ \nabla \mathbf{B}})$}
\newcommand{\minlgradbovera}{$\mathrm{min}(L_{ \nabla \mathbf{B}})/a$}

\newcommand{\minlgradgradbovera}{$\mathrm{min}(L_{ \nabla \nabla \mathbf{B}})/a$}
\newcommand{\lgradgradb}{$L_{ \nabla \nabla \mathbf{B}}$}
\newcommand{\minlgradgradb}{$\mathrm{min}(L_{ \nabla \nabla \mathbf{B}})$}
\newcommand{\mindcs}{$\mathrm{min}(d_{cs})$}
\newcommand{\mindcsovera}{$\mathrm{min}(d_{cs})/a$}
\newcommand{\mindccovera}{$\mathrm{min}(d_{cc})/a$}
\newcommand{\mindcc}{$\mathrm{min}(d_{cc})$}
\newcommand{\bdotn}{{$\mathbf{B} \cdot \hat{\mathbf{n}}$}}

\begin{document}

\preprint{AIP/123-QED}

\title{How Does The Magnetic Gradient Scale Length Influence Complexity of Filamentary Coils in Stellarators?}
\author{John Kappel}
\author{Matt Landreman}
\email{Jkappel@umd.edu}
\affiliation{Institute for Research in Electronics and Applied Physics, University of Maryland, College Park, MD}
\author{Philipp Jurašić}
\affiliation{Proxima Fusion, Flößergasse 2, 81369, Munich Germany}
\author{Sophia A Henneberg}
\affiliation{Plasma Science and Fusion Center, Massachusetts Institute of Technology, Cambridge, MA 02139, U.S.A.}

\date{\today}

\begin{abstract}

The distance between the last closed flux surface (LCFS) and the nearest electromagnetic coils is a dominating factor in the cost, size, and engineering difficulty of stellarators. The smallest magnetic gradient scale length on the LCFS - denoted \minlgradb{} - has been shown to be a good proxy for minimum coil-surface distance in optimizations of a current potential on a winding surface, such as through the REGCOIL method. However, it has not been shown the same is true for filament coils, or that the magnetic gradient scale length is an effective objective function in optimization. In this paper, we explore examples in which \minlgradb{} is correlated with the minimum coil-surface distance for filament coils. First, we analyze a subset of the single-stage-optimized equilibria from the QUASR dataset [Giuliani \textit{et al.} JPP (2024)]. We find that the majority of configurations have \minlgradb{} located near the point of closest coil-surface distance. Second, we optimize quasihelically symmetric equilibria to have improved \minlgradb{}, and optimize coils via a continuation method. We then traced alpha particles to test confinement. Finally, we compare \minlgradb{} to the minimum coil-surface distance with filament coils optimized for a set of finite beta equilibria with random boundary shapes. For all datasets, we find that \minlgradb{} is correlated with both the minimum coil-surface and coil-coil distances if sufficient coil length is allowed. Even when there is a trade-off with proxies for confinement, optimizing for improved \minlgradb{} can result in better confinement in the presence of coils, up to a point. This is because - when holding coil-coil distance constant - equilibria with lower \minlgradb{} have a larger normal field error dominated by coil-ripple causing particle loss. Both can be reduced by increasing coil-surface distance for equilibria with a high \minlgradb{}.
\end{abstract}

\maketitle

\section{INTRODUCTION}
\label{sec:INTRO}
Stellarators are a promising method to achieve magnetically confined fusion. Stellarators have a rotational transform that is predominantly generated by the external magnetic field instead of a toroidal current inside the plasma, allowing for confinement without inductive current, making it easier to achieve stability in a steady state.\cite{boozer_stellarators_2021,helanderStellaratorTokamakPlasmas2012} However, stellarators still face a few ongoing issues, such as the design and construction of coils. Compared to tokamaks, stellarator coils can be more difficult to build because they need to recreate a non-axisymmetric last closed flux surface (LCFS). This difficultly can appear in a number of ways, but one worth highlighting is that optimized stellarators often have a small minimum coil-surface distance.\cite{lion_general_2021,najmabadi_aries-cs_2008} For a deuterium-tritium reactor, the minimum coil-surface distance \mindcs{} must be at least $\sim$ 1.2 meters in order to have enough space for both the lithium breeding blanket and neutron shielding, and is a target for designs by Type One Energy, Proxima Fusion, and Thea Energy.\cite{hegnaInfinityTwoFusion2025,lionStellarisHighfieldQuasiisodynamic2025,swansonOverviewHeliosDesign2025} Improvements in the minimum coil-surface distance are a significant factor in the minimum size and cost of a reactor. Sufficient coil-surface distance is also important to allow clearances around components, since coils may move during the evacuation of the vacuum chamber, the cooling of superconducting coils, and the energization of magnets.\cite{lazerson_error_2018} Extra clearance can reduce the cost from last minute changes during construction. Improving coil-surface distance can also reduce coil-ripple.\cite{nemovCollisionlessHighEnergy2014}

Traditionally, stellarator equilibria are optimized in a two stage approach. In stage I, the shape of the LCFS is represented using a double Fourier series. The coefficients of the Fourier series are optimized to achieve desirable physics properties (such as confinement or stability). In stage II, current carrying sources (such as coils or a current potential) are optimized to match the magnetic field of the LCFS found in stage I, subject to engineering constraints relatable to the complexity of these coils. It would be advantageous to understand the relationship between the shape of an equilibrium found in stage I and the engineering constraints for optimized coils in stage II.

In a previous paper,\cite{landremanFiguresMeritStellarators2021, kappel_magnetic_2024} we introduced a magnetic gradient scale length \lgradb{} defined as follows\footnote{The factor of $\sqrt{2}$ is chosen so that for an infinite straight wire, \lgradb{} is the same as the distance to the wire.}: 
 \begin{ceqn}
\begin{equation}
        L_{\nabla \mathbf{B}}= \sqrt{2} \frac{B}{\| \nabla \mathbf{B} \|_F},
        \label{eq:LgradB}
 \end{equation}
\end{ceqn}
 where $\mathbf{B}$ is the magnetic field vector at a point on the LCFS, and $\| \mathbf{A} \|_F = \sqrt{\sum_{i,j} A_{ij}^2}$ is the Frobenius norm of a rank 2 tensor. We optimized the current potential on a winding surface using \texttt{REGCOIL}\cite{landreman_improved_2017} and showed that for more than 40 stellarator designs, \minlgradb{} was an effective proxy for the coil-surface distance. Since the original paper, \minlgradb{} had been used as an objective function in limited contexts,\cite{helander_optimised_2024,cadenaConStellarationDatasetQIlike2025a,hegnaInfinityTwoFusion2025,schuett_optimization_2025,goodmanQuasiIsodynamicStellaratorsLow2024} but a systematic study has yet to be done to see to what extent using \minlgradb{} as an objective function improves coils. In other words, does optimizing \minlgradb{} in stage I result in improved coils in stage II?
 
In addition, our previous paper\cite{kappel_magnetic_2024} comparing \minlgradb{} to coil-surface distance used a current potential on a winding surface, which is a simplified model for coils. It is not guaranteed that \minlgradb{} would also be an effective proxy with a more complex but realistic model, such as filamentary coils. Comparing \minlgradb{} to \mindcs{} is more complicated for filamentary coils than for a current potential for two reasons. First, in the winding surface method, we generated the winding surface by taking the LCFS and extending it everywhere by a uniform distance normal to the LCFS. In contrast, there is no single coil-surface distance for filamentary coil methods, unless the coils are constrained to lie on the LCFS. Second, there is one natural coil complexity metric associated with the winding surface method. This was the maximum surface current density, which we had written as $\|\mathbf{K}\|_\infty$. When comparing coil complexity across equilibria, we had held $\|\mathbf{K}\|_\infty$ constant. However, when using filamentary coils, there are many coil complexity metrics, including coil length $L$, minimum coil-coil distance \mindcc{}, and maximum coil curvature $\max(\kappa)$. It may not be possible or fair to hold them all fixed between configurations. As a result, it is not clear which quantities to hold constant when comparing equilibria.

In this paper, we explore the validity \minlgradb{} as a predictor of coil-surface distance when optimizing with filamentary coils. We find that improving \minlgradb{} generally improves both the minimum coil-surface distance and minimum coil-coil distance if sufficient coil length is allowed. Even when there is a trade-off with proxies for confinement, we find that optimizing for improved \minlgradb{} can result in better confinement in the presence of coils, up to a point. This is because - when holding engineering constraints such as coil-coil distance constant - equilibria with lower \minlgradb{} have a larger normal field error dominated by coil-ripple. Reducing coil-ripple by increasing coil-surface distance for equilibria with a high \minlgradb{} can improve $\alpha$ particle loss to overcome the trade-off with quasi-symmetric error during stage I optimization. 

The rest of this paper is structured as follows. In Section \ref{sec:QUASR}, we look at the coil-surface distance for the QUAsisymmetric Stellarator Repository (QUASR) equilibria and coils, previously optimized using single stage optimization.\cite{giuliani_direct_2023} In Section \ref{sec:QH}, we optimize quasihelisymmetric plasmas for better \lgradb{} and show that the coil-surface distance and $\alpha$ particle confinement can be improved. In Section \ref{sec:RAND}, we optimize filamentary coils for a set of finite $\beta$ equilibria. These equilibria have random boundary shapes and number of field periods equal to 2 ($N_{fp} = 2$). With the optimized coils, we compare \minlgradb{} to \mindcs{}. We find that there is a correlation between \minlgradb{} and \mindcs{}, though not as strong as for the dataset in section \ref{sec:QH}.

\section{coil-surface Distance for Single Stage Optimized QUASR Equilibria} \label{sec:QUASR}
To assess the validity of using \minlgradb{} for filamentary coils as a proxy for \mindcs{}, we started by analyzing data from the existing Quasi-Symmetric Stellarator Repository (QUASR),\cite{giuliani_quasr_2024} which is a database of over 300,000 optimized vacuum stellarators with quasi-axisymmetric (QA) or quasi-helical (QH) symmetry. These configurations were generated using single stage optimization: as opposed to the stage I and stage II optimization method described in section \ref{sec:INTRO}, the coils and plasma were optimized in tandem. All configurations have an average major radius $R_0$ of 1 meter. 
\begin{figure}[tbp]
     \centering
     \hfill
     \includegraphics[width = \linewidth]{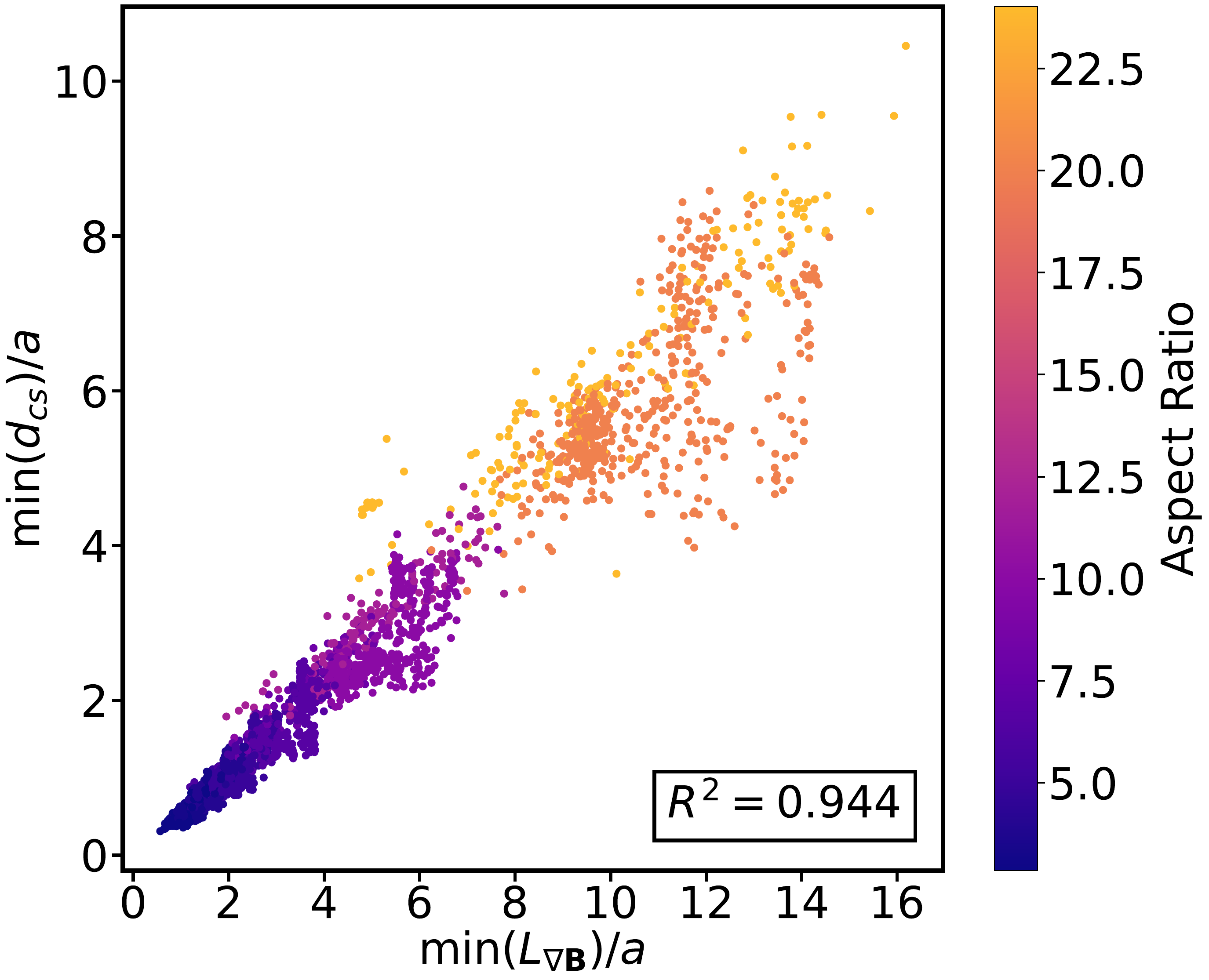}
     \caption{Properties of plasma equilibria and associated coil sets for a subset of 3027 QUASR configurations.  While initially the correlation between \minlgradbovera{} and \mindcs{} seems strong, some of the correlation can be explained by the variation in the aspect ratio. Generally, equilibria with larger aspect ratio have larger \mindcsovera{}.}
    \label{fig:QUASR_lgradb_over_a_full}
\end{figure}
\begin{figure}[tbp]
     \centering
     \includegraphics[width= \linewidth]{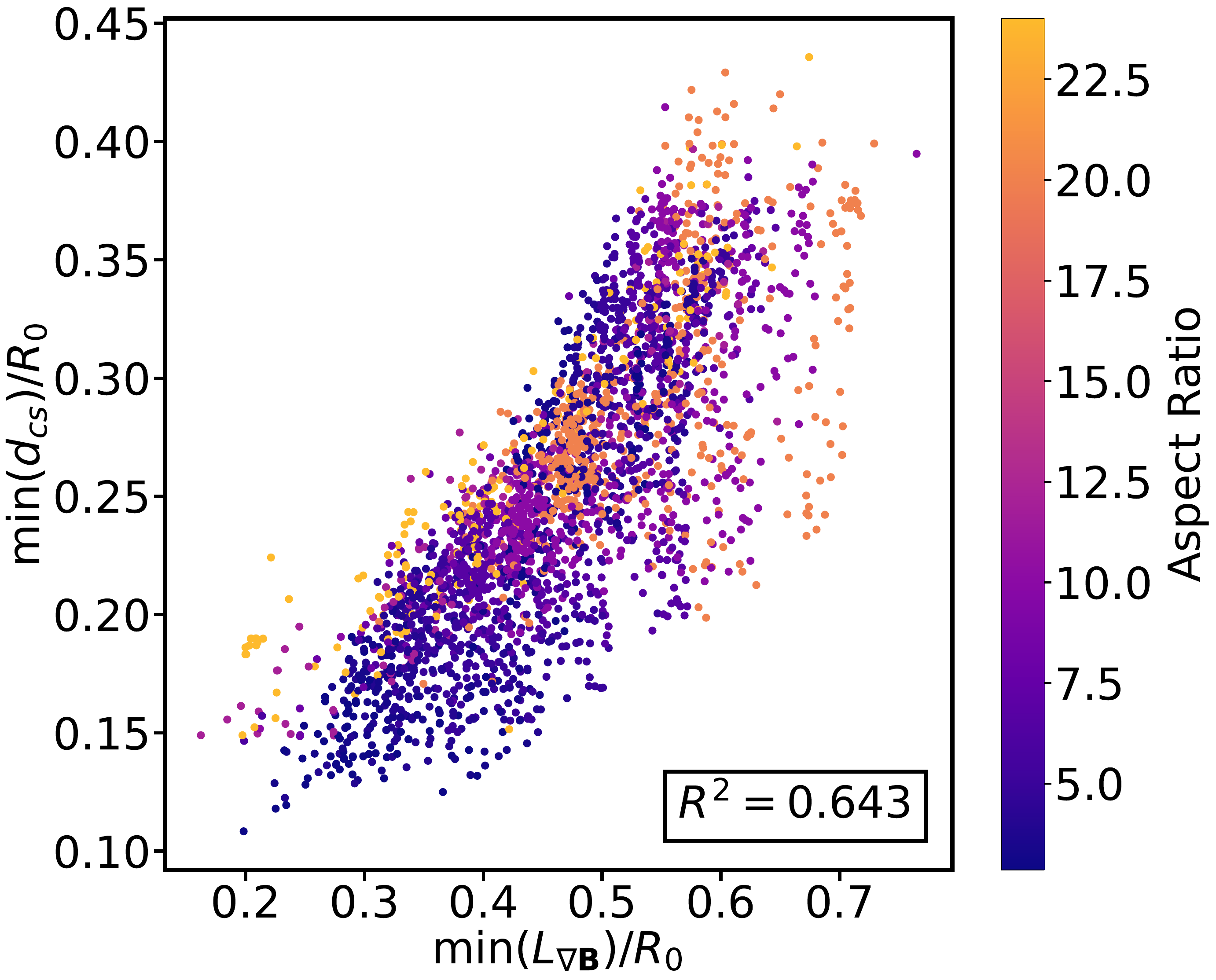}
     \hfill
     \caption{Properties of plasma equilibria and associated coil sets for a subset of 3027 QUASR configurations. This is the same dataset shown in figure \ref{fig:QUASR_lgradb_over_a_full} normalized by $R_0$ instead of $a$. When normalized by major radius,  \minlgradb{} remains correlated with \mindcs{}.}
    \label{fig:QUASR_lgradb_full}
\end{figure} 
A random subset of 6000 equilibria and associated coils were chosen from the QUASR database with varying aspect ratio, quasisymmetry type (QA vs QH), and number of field periods. We only evaluate the magnetic field on the LCFS because that is where $\| \nabla \mathbf{B} \|_F$ is the largest in vacuum. A proof of this fact is shown in Appendix \ref{app:gradB}. These configurations were chosen to  have same number of total coils. This is because the optimization includes a pairwise coil-coil distance penalty in the objective function,\cite{giuliani2024direct,giuliani2025comprehensive} which "pushes" the coils away from each other and as a consequence closer to the plasma. In order to have a clean comparison between coil penalty terms and following work by Jurašić, \cite{jurasicNovelApproachCombined2024} the dataset only includes configurations with 12 total coils. This includes $n_{fp}$ = 6, $n_{fp}$ = 3 and $n_{fp}$ = 2 equilibria,  but not $n_{fp}$ = 4, since QUASR only includes configurations with an even number of coils per field period.  

These 6000 coil sets were then filtered so that $I_{min}/I_{max} > 0.5$, where $I_{max}$ is the largest current for a coil set and $I_{max}$ is the smallest. As motivation for this constraint, consider the extreme case where $I_{min} = 0$. The magnetic field is then independent of the location of the zero-current coil. In this case, the zero-current coil can be moved closer to the plasma, decreasing \mindcs{} without changing \minlgradb{}, making \minlgradb{} a poor proxy. The same is true if two adjacent coils have currents flowing in opposite directions, since one could place the two coils an infinitesimal distance away from each other, and they would behave like a zero-current coil. Ideally, $I_{min}/I_{max}=1$ to avoid this concern, but the weaker constraint $I_{min}/I_{max}> 0.5$ is chosen so that a larger number of equilibria could be used. We found this was the minimum cutoff before correlation worsened. These coil sets were filtered so that all currents flowed in the same physical direction.  After this filtering, we had 3027 equilibria with associated coil sets.

For each configuration, the magnetic field was calculated by evaluating the Biot-Savart law with \texttt{SIMSOPT}.\cite{noauthor_hiddensymmetriessimsopt_2025} The number of points used to calculate the distances between the coils and surface are higher than in the original QUASR dataset: 300 points  were used along each coil, and 128 by 128 points per half field period were used on each surface. 
Most of the configurations in the dataset have \mindcc{} within a small interval: for 90.3\% of the data, \mindcc{} is between 0.091 m and 0.11 m. 

In figure \ref{fig:QUASR_lgradb_over_a_full}, we compare \minlgradbovera{} to \mindcsovera{} for this random subset of quasisymmetric vacuum stellarators from QUASR. 
A high correlation between these quantities can be seen, with a coefficient of determination $R^2$ of 0.944.
The configurations vary widely in minor radius (between 0.0416 and 0.3519 m) and therefore also in aspect ratio (between 2.8091 and 24.01), indicated by the color.
It can be seen that both \minlgradbovera{} to \mindcsovera{} are correlated with the aspect ratio.

These trends with aspect ratio can be understood by considering a purely toroidal axisymmetric field. In this case it can be shown that \lgradb{} is equal to the local major radius $R$ on the surface, and that \minlgradb{} is on the inboard side and equals $R_0 - a$. Therefore \minlgradbovera{} = $R_0/a-1$. This field can be generated by an infinite straight wire along the $z$ axis, meaning that \mindcsovera{} = \minlgradbovera{}. In other words, for equilibria similar to an axisymmetric field, we expect both \minlgradbovera{} and \mindcsovera{} to scale linearly with aspect ratio, which is what we see in figure \ref{fig:QUASR_lgradb_over_a_full}.

Since some of the correlation in figure \ref{fig:QUASR_lgradb_over_a_full} can be explained by the variation in aspect ratio, in figure \ref{fig:QUASR_lgradb_full} we have plotted \minlgradb{} and \mindcs{} normalizing by the average major radius $R_0$ instead of minor radius. 
This $R_0$ normalization eliminates the overall trends with aspect ratio, so the correlation of \minlgradb{} with \mindcs{} can be assessed independent of the aspect ratio.
While \minlgradb{} and \mindcs{} remain correlated, $R^2$ is more moderate, 0.644.
Normalizing by the minor radius and major radius both have validity, and so it is unclear which normalization is the "right" one. Either way, we can conclude that \minlgradb{} is predictive of \mindcs{}.

\begin{figure}[tbp]
     \centering
     \includegraphics[width= \linewidth]{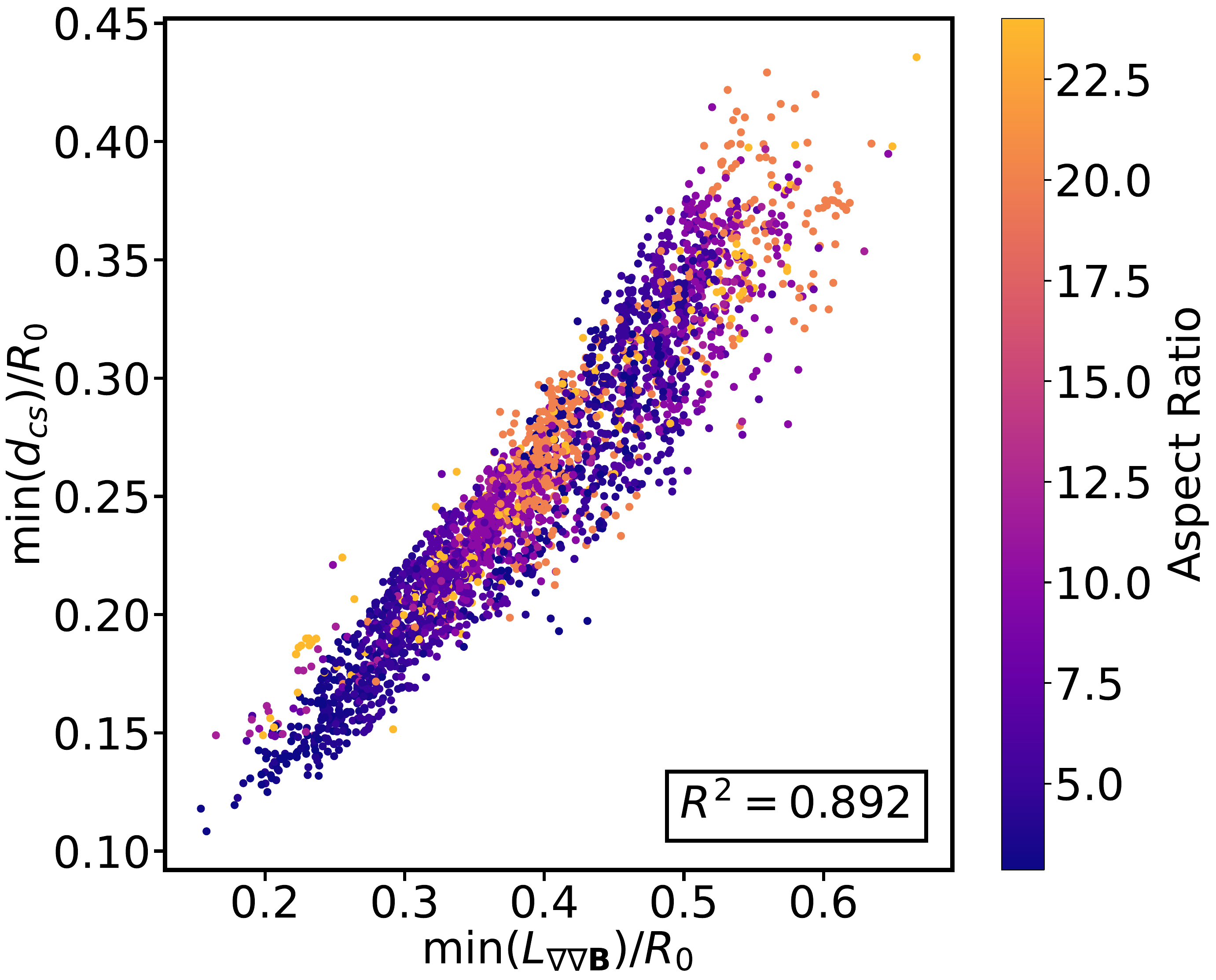}
     \hfill
     \caption{Properties of plasma equilibria and associated coil sets for a subset of 3027 QUASR configurations. This is the same dataset shown in figure \ref{fig:QUASR_lgradb_full}. When the length scale on the X axis has been changed from equation \ref{eq:LgradB} to equation \ref{eq:LgradgradB}, a stronger correlation is found.}
    \label{fig:QUASR_lgradgradb_full}
\end{figure}

In our previous paper, we had speculated that looking at a scale length related to the norm of $\nabla \nabla \mathbf{B}$ may also be an effective proxy for coil-surface distance. Norms of $ \nabla \mathbf{B}$  have only information from the first gradient, making it a very localized scale length. Norms of $\nabla \nabla \mathbf{B}$ have information from the second gradient, meaning it reflects a larger neighborhood around a point. \cite{kappel_magnetic_2024} Following Landreman (2021),\cite{landremanFiguresMeritStellarators2021} we calculated the following metric:

\begin{ceqn}
\begin{equation}
    L_{\nabla\nabla \mathbf{B}}  = \sqrt{\frac{4B}{\mathbf{\|\nabla\nabla\mathbf{B}}\|_F}},
\label{eq:LgradgradB}
\end{equation}
\end{ceqn}
 where $\| \mathbf{A} \|_F = \sqrt{A \threedots A} = \sqrt{\sum_{i,j,k} A_{ijk}^2}$ is the Frobenius norm of a rank-3 tensor. Like \lgradb{}, \lgradgradb{} is normalized so that when evaluated for the magnetic field of an infinite straight wire, it is equal to the distance to the wire. In figure \ref{fig:QUASR_lgradgradb_full}, we compare \minlgradgradb{} with \mindcs{}. We find a strong correlation between these two quantities, stronger in fact than between \mindcs{} and \minlgradb{}.

\begin{figure}[tbp]
     \centering
     \includegraphics[width= \linewidth]{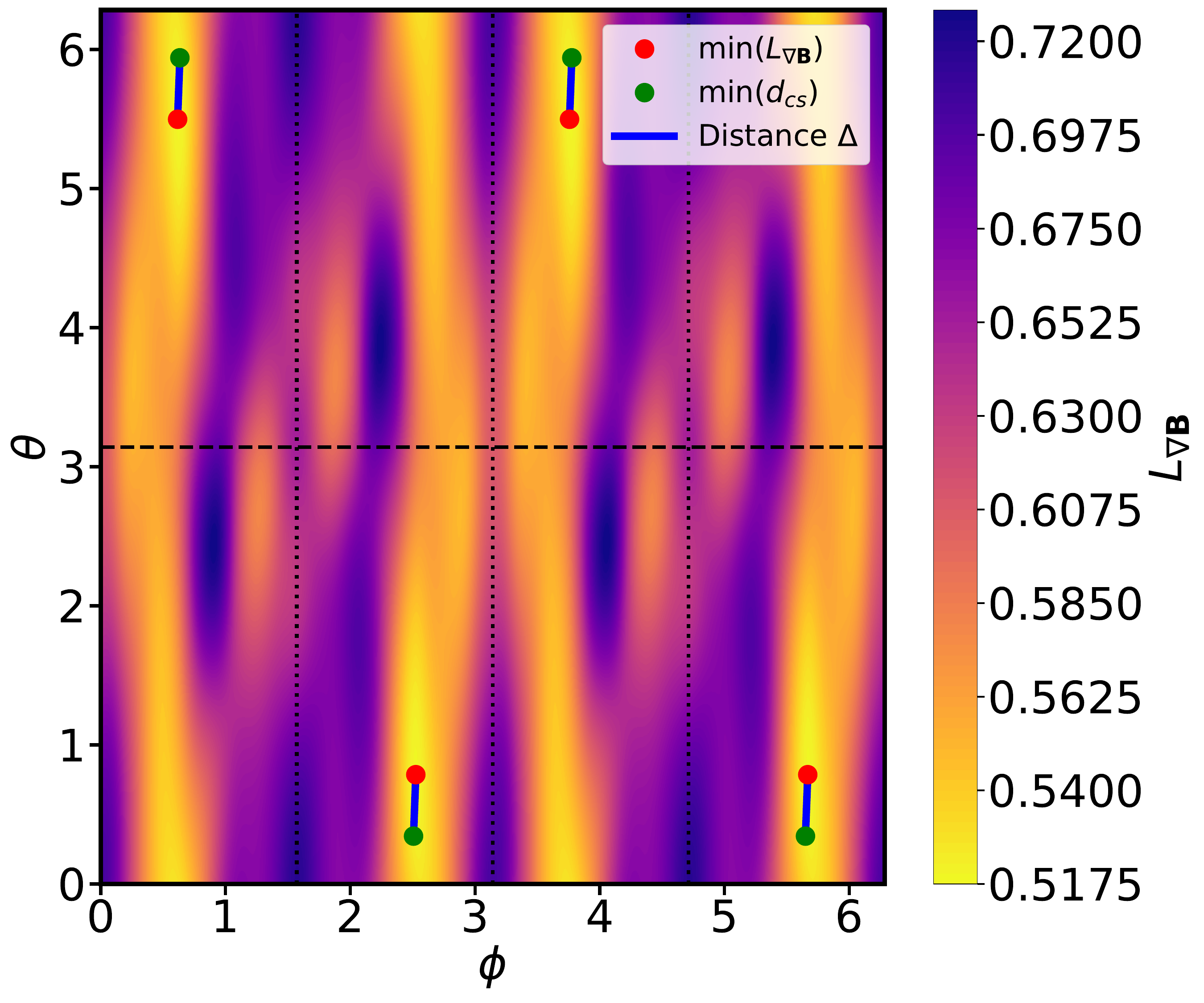}
     \hfill
     \caption{This plot depicts the LCFS for an example equilibrium from the QUASR dataset in terms of $\theta$ and $\phi$. The colorbar represents \lgradb{}, where the smallest values are in yellow. Black dotted lines depict planes of symmetry. In red, the locations of \minlgradb{} are shown. In green, the locations of the smallest $d_{cs}$ are shown. The closest distance between \minlgradb{} and \mindcs{} is  measured by $\Delta$ and is shown by the blue lines. For this example, $\Delta$ is 0.1.}
    \label{fig:example_delta}
\end{figure}

\begin{figure}[tbp]
     \centering
     \includegraphics[width= \linewidth]{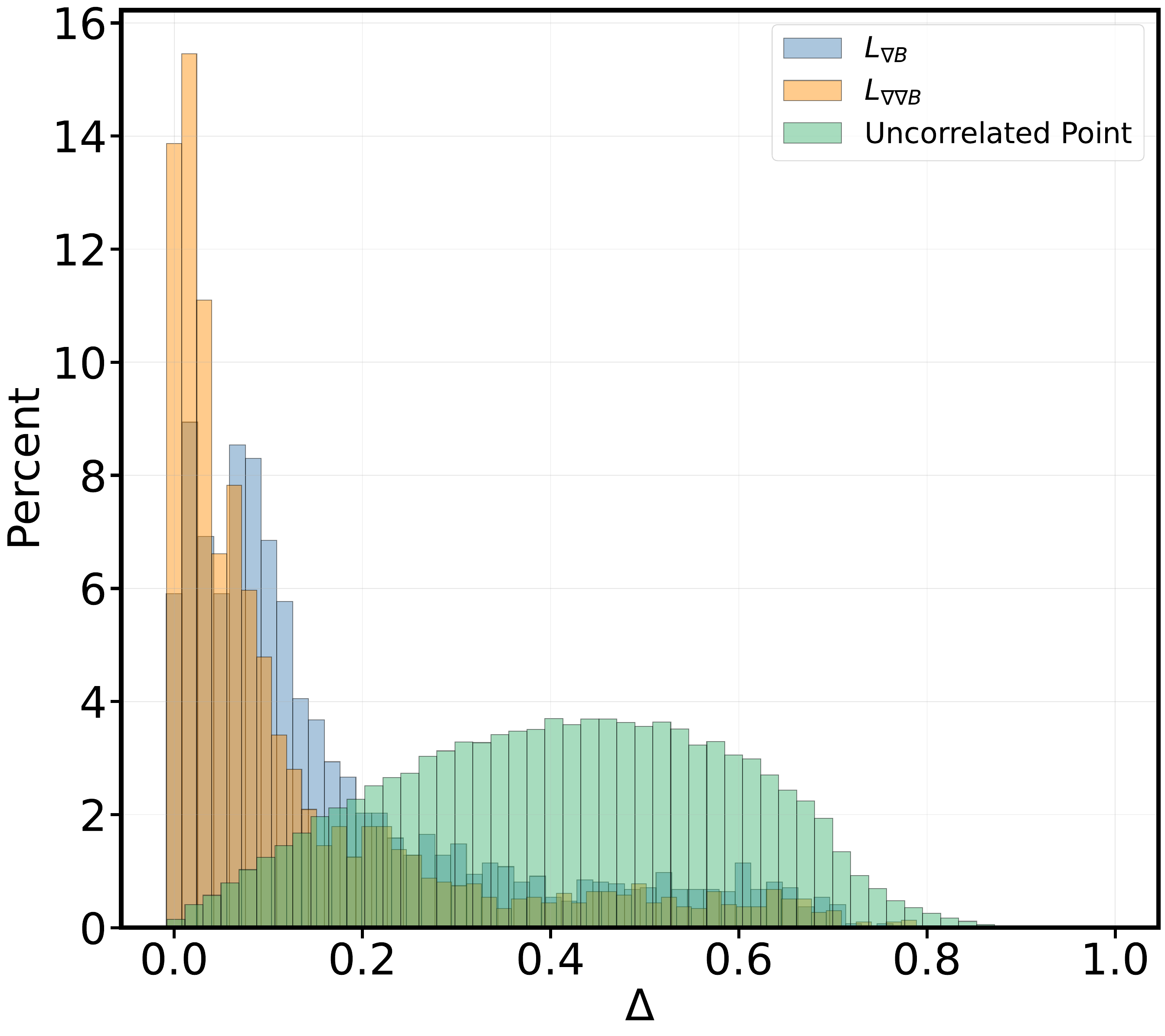}
     \hfill
     \caption{A histogram of distances from points of \mindcs{} to points of \minlgradb{} or \minlgradgradb{} for 3027 equilibria in QUASR. $\Delta$ measures the distance between the locations of the smallest coil-surface distance and the lowest \lgradb{} or \lgradgradb{}. For \lgradb{}, 44.5\% of equilibria have $\Delta$ less than 0.1. For \lgradgradb{}. 62.7\% of equilibria have $\Delta$ less than 0.1.}
    \label{fig:hist_lgradB}
\end{figure}
 
In addition, in most configurations, the coils are closest to the plasma near the point of smallest \lgradb{} and smallest \lgradgradb{}. To show this, we performed the following analysis. We first found the locations on the LCFS where the coil-surface distance was smallest. We then found the locations on the LCFS where \lgradb{} was smallest. For both quantities, the minimum repeats $2n_{fp}$ times due to both field period and stellarator symmetry. We then calculated an effective distance in one field period of the $\theta-\phi$ plane, $\Delta =  \sqrt{\delta\theta ^2+ (n_{fp}\delta\phi)^2}/(\sqrt{2}\pi)$, where $\delta\theta$ and $\delta \phi$ are the differences in $\theta$ and $\phi$ between the minima of \lgradb{} and $d_{cs}$, accounting for all the symmetric copies. This formula is normalized to account for both field period and stellarator symmetry, so that the farthest possible points - on opposite sides poloidally and half a field period away toroidally - would have a $\Delta$ of 1. We show an example in figure \ref{fig:example_delta} for one such equilibrium. We then repeated this analysis for \minlgradgradb{}.  For a point of comparison, we also evaluated the distribution of $\Delta$ for two points on the LCFS that are completely uncorrelated. In this test, we randomly chose  $\theta$ and $\phi $ between 0 and $2\pi$ from a uniform distribution. 

In figure \ref{fig:hist_lgradB}, we show a histogram depicting the distribution of $\Delta$ for all 3027 equilibria and coils for \minlgradb{}, \minlgradgradb{}, and the uncorrelated points. The mean $\Delta$ for \minlgradb{} and \minlgradgradb{} are 0.18 and 0.14, respectively. These are significantly lower than for the uncorrelated points, which have a $\Delta$ of 0.43. In addition, let us measure the percent of equilibria with a $\Delta$ below 0.1 (this is the $\Delta$ shown in figure \ref{fig:example_delta}
). For \minlgradb{}, this is 44.5\%; for \minlgradgradb{} this is 62.7\%; for an uncorrelated measurement this is 3.05\%. These metrics show that $\Delta$ is significantly smaller between \mindcs{} and \minlgradb{} than for uncorrelated points, and smaller still between \mindcs{} and \minlgradb{}. This shows that for most equilibria, \minlgradb{} and especially \minlgradgradb{} tend to be located at the same point on the LCFS as the point of minimum coil-surface distance. This spatial correlation strengthens the case that coil-plasma distance is fundamentally limited by scale lengths in the magnetic field.

\section{Optimizing QH Equilibria using \texorpdfstring{\minlgradb{}}{LgradB} Results in Coils with Favorable Engineering Quantities}
\label{sec:QH}

In this section, we will examine the relationship between \minlgradb{} and \mindcs{} for an additional dataset. Rather than just seeing if there is correlation between these two quantities for preexisting equilibria, we intentionally manipulate \minlgradb{} to see how it affects \mindcs{}. Ideally, we can obtain an improved value for \minlgradb{} without sacrificing other desirable physics properties like confinement to achieve it. While in the previous section we showed strong correlation between \mindcs{} and \minlgradgradb{}, \minlgradgradb{} requires a higher spectral resolution in order to resolve, making it less promising as an optimization metric. Therefore, we chose to only look at manipulating \minlgradb{} in this section. To manipulate \minlgradb{} we first perform stage I plasma optimization while varying the minimum threshold of this quantity. We then perform stage II coil optimization for a variety of coil lengths.
Each stage is described in more detail in the following sections.

\subsection{Stage I Optimization To Improve \texorpdfstring{\minlgradb{}}{min LgradB}}
\label{sec:QH_stageI}

\begin{figure}[tbp]
     \centering
     \includegraphics[width= \linewidth]{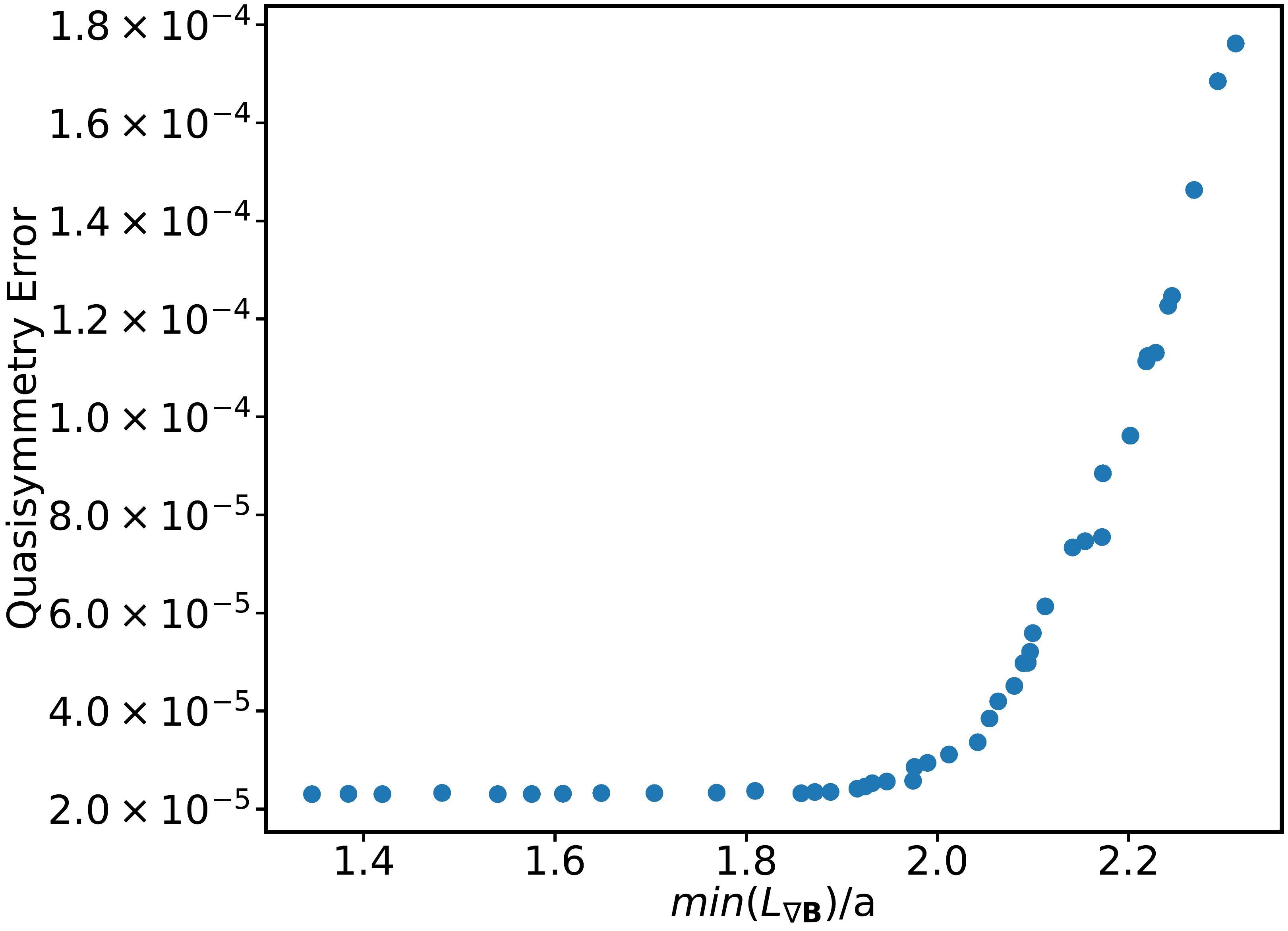}
     \hfill
     \caption{Stage 1 optimized QH plasma equilibria in section \ref{sec:QH_stageI}. Each dot represents a different equilibrium with a different target \lgradb{}. There is a tradeoff between the minimum \lgradb{} and the quasisymmetry error.}
    \label{fig:stage_1}
\end{figure}

\begin{figure}[tbp]
     \centering
     \includegraphics[width= \linewidth]{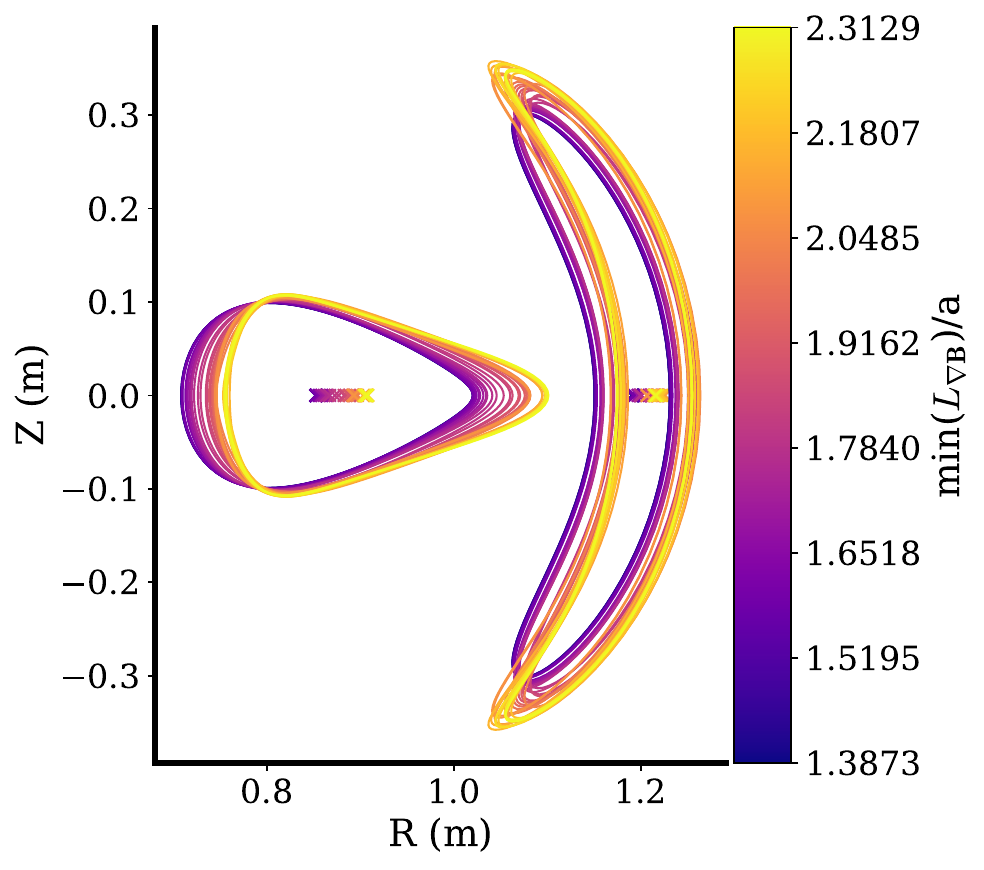}
     \hfill
     \caption{Cross sections of constant $\phi$ for the family of optimized QH configurations in section \ref{sec:QH_stageI}. Shown on the right is the $\phi = 0$ cross section (often referred to as the bean cross section). Shown on the left is the $\phi = \pi/4$ cross section (often referred to as the triangular cross section).}
    \label{fig:bean_and_triangle}
\end{figure}

\begin{figure}[tbp]
     \centering
     \includegraphics[width= \linewidth]{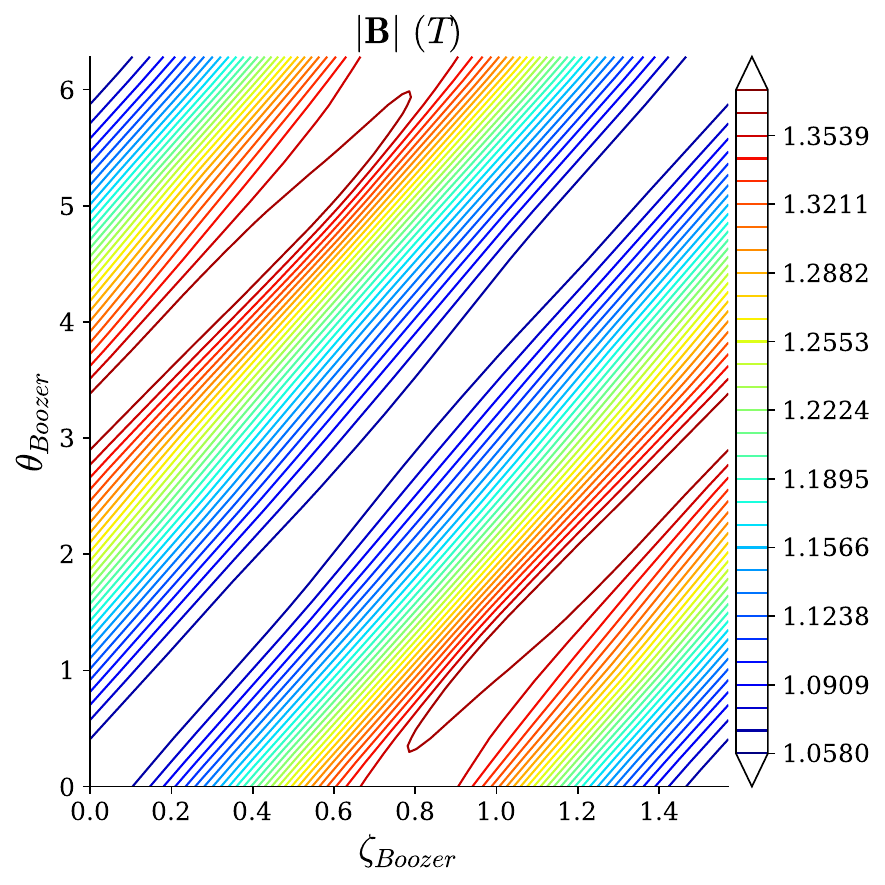}
     \hfill
     \caption{The magnitude of the magnetic field on the fixed  boundary of a QH stellarator with \minlgradbovera{} = 1.35 after stage I optimization. This is the equilibrium with the smallest \minlgradbovera{} from section \ref{sec:QH_stageI}. The field is plotted in Boozer coordinates, which means that if the field was quasisymmetric, the contours of constant field strength should be straight lines.}
     \label{fig:stage_1_boozer_low_lgradB}
\end{figure}

\begin{figure}[tbp]
     \centering
     \includegraphics[width= \linewidth]{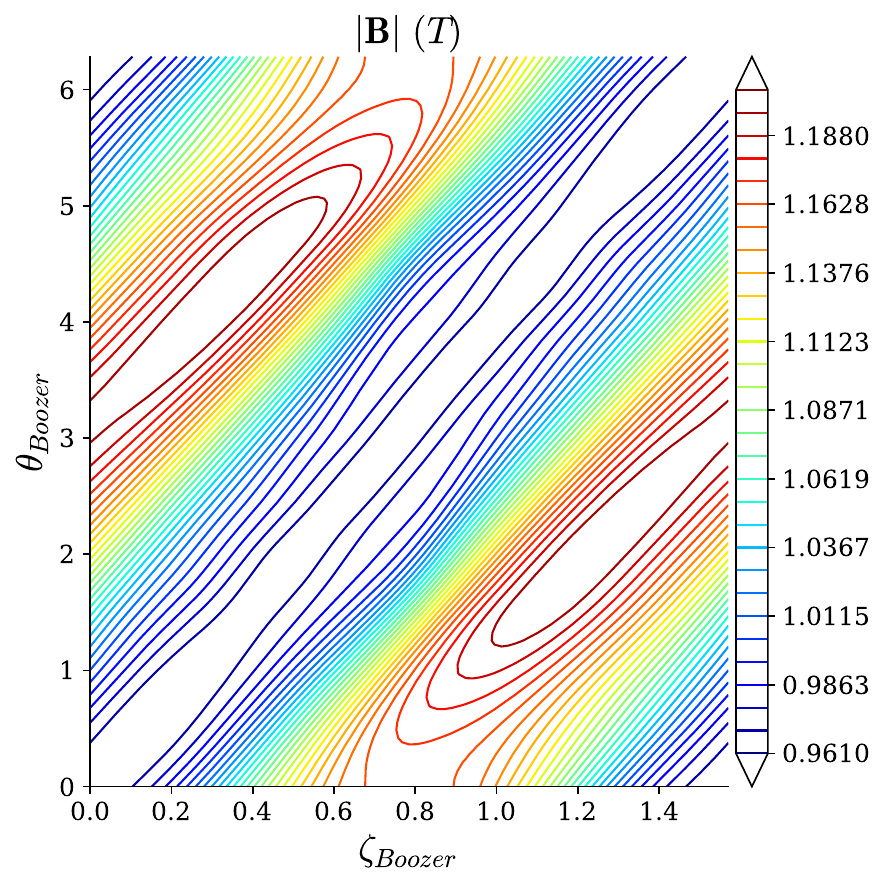}
     \hfill
     \caption{The magnitude of the magnetic field on the fixed  boundary of a QH stellarator with \minlgradbovera{} = 2.31 after stage I optimization. This is the equilibrium with the largest \minlgradbovera{} from \ref{sec:QH_stageI}. The field is plotted in Boozer coordinates, which means that if the field was quasisymmetric, the contours of constant field strength should be straight lines. This equilibrium has worse (higher) quasisymmetry error compared to the one in figure \ref{fig:stage_1_boozer_low_lgradB}.}
     \label{fig:stage_1_boozer_high_lgradB}
\end{figure}

In this subsection, we present a method for optimizing stellarators to improve \minlgradb{}. The resulting family of optimized QH stellarators are similar to the Helically Symmetric eXperiment (HSX).\cite{andersonHelicallySymmetricExperiment1995} Both this configuration family and HSX are $N_{fp} = 4$ QH vacuum equilibria, have a magnetic well, and have $\iota$ slightly above 1 to avoid low order rationals. We optimized a using an initial condition of a rotating ellipse with $N_{fp}$ = 4 in vacuum using the following objective function:

\begingroup
\setlength{\mathindent}{0pt}
\begin{subequations}
\begin{gather}
f  = \left(  \frac{R_{0}/a - 8}{8} \right)^2 + \label{eq:aspect}
\\\left(\frac{\mathrm{\iota(0) - 1.03}}{1.03} \right)^2 + \label{eq:iota_core}\\
\left( \frac{\mathrm{max( \iota{(1)-1.085,0})}}{1.085} \right)^2 +
\label{eq:iota_edge}\\
\int \frac{d^3 x}{B^6} [(4-\iota)\mathbf{B} \times \nabla  B \cdot \nabla  \psi - G \mathbf{B} \cdot \nabla B]^2  +
\label{eq:QS}\\
 \int_{0}^1 d \rho \;\mathrm{max} \left( \frac{1}{V} \frac{d^2V}{ds^2}  + 0.013, 0 \right)^2  +
\label{eq:magnetic_well}\\
\biggl< \mathrm{max} \left( \frac{a}{B} \|\nabla \mathbf{B} \|_{F} - C, 0 \right)^2 \biggr>_{\rho=1}, \label{eq:lgradB_opt}
\end{gather}
\end{subequations}
\endgroup
where $R_0$ is the average major radius and $a$ is the average minor radius, $\iota(0)$ is the rotational transform on the axis, $\iota(1)$ is the rotational transform on the LCFS, $G$ is $\mu_0/(2\pi)$ times the current linking the plasma poloidally, $V$ is the volume of the flux surface,  $s=\rho^2$ is the normalized toroidal flux, and $\langle \cdot \rangle_{\rho=1}$ is the flux surface average on the LCFS.
No weights multiplying each term were necessary.
The term \ref{eq:aspect} targets an aspect ratio of 8. The aspect ratio is held constant so we can compare improvements in \mindcs{} independent to any changes in aspect ratio. The terms \ref{eq:iota_core} and \ref{eq:iota_edge} ensure the equilibrium does not have a low-order rational rotational transform $\iota$ on any flux surface. The term \ref{eq:QS} targets quasisymmetry error. The term \ref{eq:magnetic_well} ensures the equilibrium has a magnetic well. The term \ref{eq:lgradB_opt} penalizes small \minlgradb{}.  Using this objective, we scanned through \minlgradb{} by changing $C$ in the term \ref{eq:lgradB_opt}, resulting in a Pareto front of 42 equilibria. Optimizations were performed using \texttt{DESC}.\cite{dudtDESCStellaratorEquilibrium2020} These results are shown in figure \ref{fig:stage_1} and show the tradeoff between \minlgradbovera{} and quasisymmetry error. The bean and triangle cross sections are shown in figure \ref{fig:bean_and_triangle}. The shapes are qualitatively similar between the equilibria. Improving \minlgradb{} actually increases the curvature of the tips of the bean cross section, meaning the surface curvature is not reflective of the $B$ scale length. We show the magnitude of the field strength in Boozer coordinates for the equilibria with lowest and highest \minlgradb{} in figures \ref{fig:stage_1_boozer_low_lgradB} and \ref{fig:stage_1_boozer_high_lgradB} respectively. 


\subsection{Stage II Optimization to Design Coils with Good Engineering Constraints}
\label{sec:QH_coils}

\begin{figure}[tbp]
     \centering
     \includegraphics[width= \linewidth]{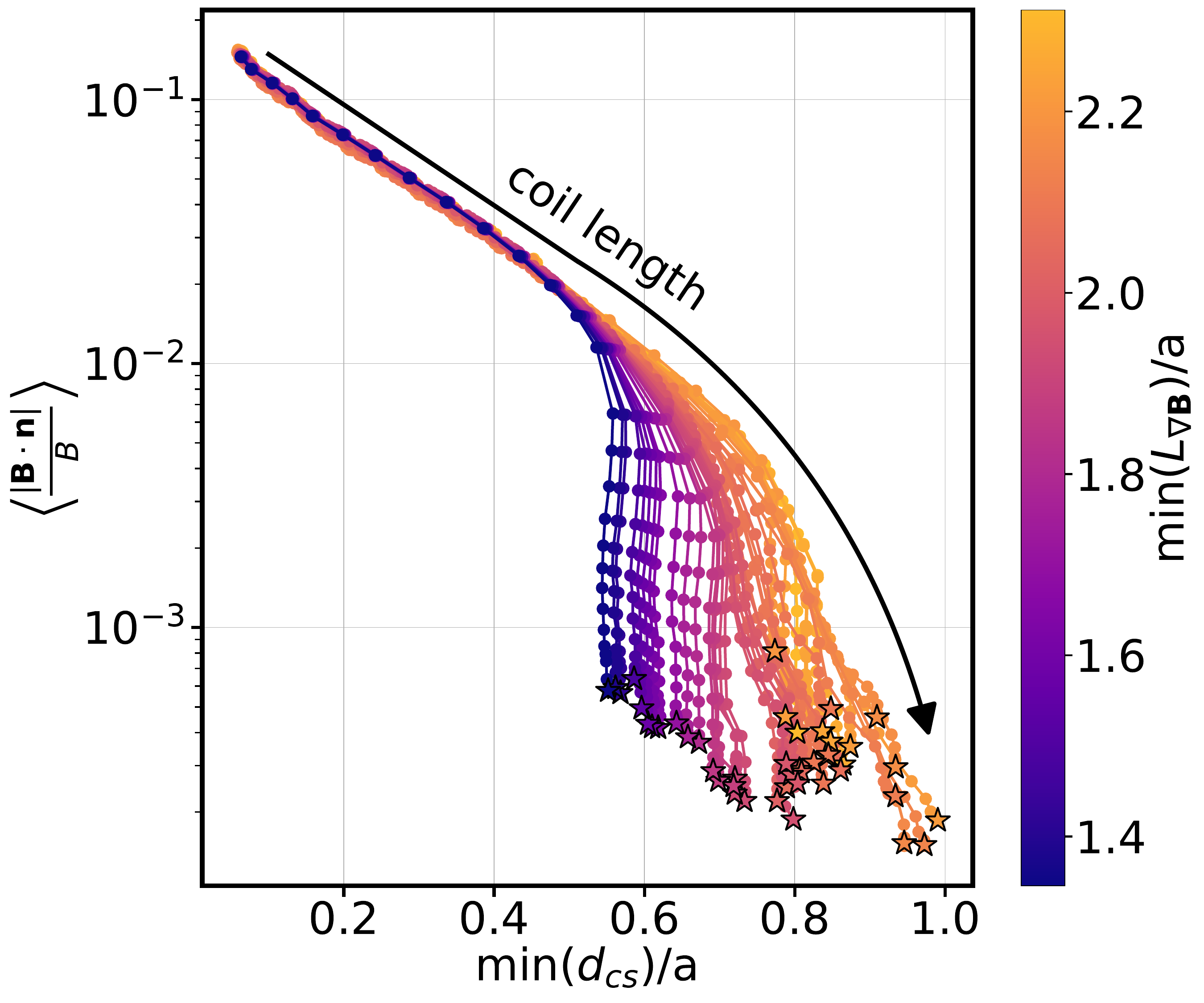}
     \hfill
     \caption{The data related to a series of stage II coil optimizations are plotted. Each colored curve represents a different  member in a family of related QH equilibria, each with a different minimum \lgradb{}$/a$. This is the data set described in detail in section \ref{sec:QH}. Starting in the top left corner and moving down right, each dot represents a coil set optimized by a subsequent round of the continuation method, where the objective function is modified to increase coil length. The objective function is modified and the previous solution is used as a warm start for the next dot. Each star represents the final coil set optimized for a target of 3.6 meters per coil. There is a trend that equilibria with better minimum \lgradb{}$/a$ have improved minimum coil-surface distance (on the X axis) and average absolute normal field error (on the Y axis).}
    \label{fig:QH_stage2_cs}
\end{figure}

\begin{figure}[tbp]
     \centering
     \includegraphics[width= \linewidth]{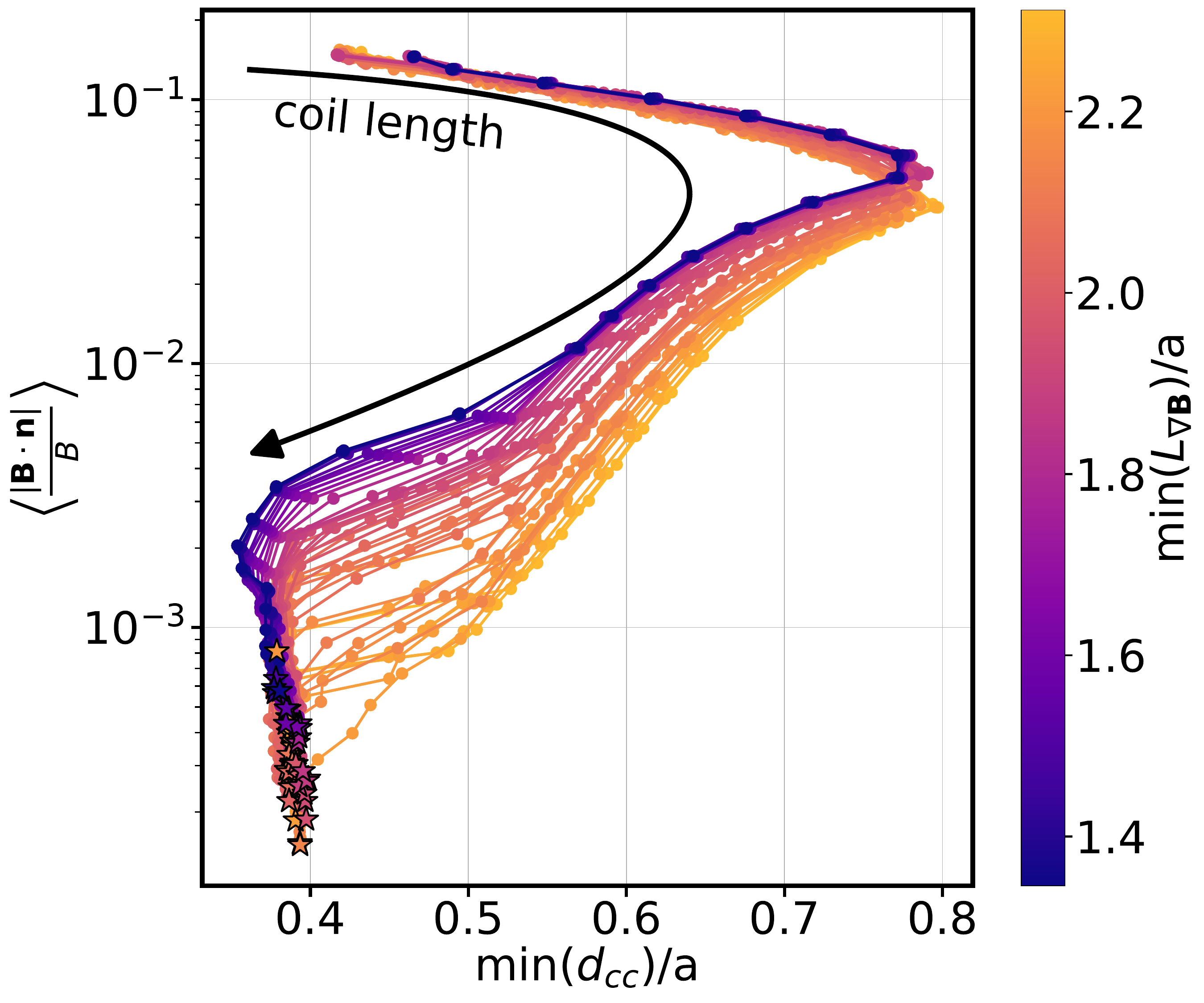}
     \hfill
     \caption{The data related to a series of stage II coil optimizations are plotted. These are the same coils shown in figure \ref{fig:QH_stage2_cs}, but the X axis is changed to show how the coil-coil distance changes as coil length increases. \mindccovera{} is small, then gets larger, then gets smaller again. There is a trend that equilibria with better minimum \lgradb{}$/a$ have improved minimum plasma coil-coil distance (on the X axis) and average absolute normal field error (on the Y axis).}
    \label{fig:QH_stage2_cc}
\end{figure}

For each equilibrium resulting from the stage I optimization in the previous subsection, we then optimize a series of coils with varying length. We used a continuation method in which we modified the objective function after each round of optimization, and used the previous round of coils as a warm start for the next round of optimization. For each round $i$, the objective function was
\begin{subequations}
\begin{gather}
f_{i} = w_{\mathbf{B \cdot \hat{n}}} f_{\mathbf{B \cdot \hat{n}}} + w_{cc} f_{cc} + w_l f_{l,i} + f_{cs,i} \\
f_{\mathbf{B\cdot \hat{n}}} = \frac{\iint_S \; ds (\mathbf{B_{total} \cdot \hat{n} } )^2}{\iint_S ds B^2} \\
f_{cc} = \mathrm{min}((d_{cc}-d_{cc}^*),0)^2 \\ 
f_{l,i} =  \sum_{n} (l_n-l_{i}^*)^2 \label{eq:l} \\
f_{cs,i} = \mathrm{min}((d_{cs}-d_{cs,i}^*),0)^2 \label{eq:d_cs} \\ 
d_{cs,i}^* = d_{cs,i-1}  \label{eq:d_csi}\\
\mathbf{B_{total}} =  \mathbf{B_{coils}}  + \mathbf{B_{plasma}}
\end{gather}
\end{subequations}
where each $w$ is the weight on each objective, $\mathbf{B_{total} \cdot \hat{n}}$ is the normal field error, $\iint_S ds$ is a surface integral on the LCFS, $\mathbf{B_{coils}}$ is the magnetic field vector generated by the coils, $\mathbf{B_{plasma}}$ is the magnetic field vector generated by plasma current (which is 0 for vacuum fields), $\mathbf{\hat{n}}$ is the unit vector normal to the LCFS, $d_{cc}$ is the minimum coil-coil distance, $d_{cc}^*$ is the target minimum coil-coil distance, $l_n$ is the length for each coil, and $l_i^*$ is the target length for each coil. In each round, the target length $l_i^*$ increases. This method allows us to scan the length, so coils of differing length can be compared. We optimized with six coils per half field period with fixed current using \texttt{SIMSOPT}. For each equilibrium, coils were optimized with 20 length targets ranging from 1.11 meters per coil to 3.60 meters per coil. The minimum coil-coil distance threshold $d_{cc}^*$ was chosen to be 0.05 meters, which is equivalent to $d_{cc}^* / a = 0.4$. Coils were initialized as circles lying normal to the magnetic axis. Equations \ref{eq:d_cs} \& \ref{eq:d_csi} also act as a soft constraint on the coil-surface distance. We add this term to ensure that increasing coil length is associated with increased coil-surface distance, rather than letting the coils get closer to the plasma by forming helical loops. The results of these optimizations are shown in figures \ref{fig:QH_stage2_cs} and \ref{fig:QH_stage2_cc}, showing coil-surface distance and coil-coil distance respectively. Note that since all equilibria have the same aspect ratio, the results will be the same if we normalize lengths to the average minor radius or average major radius. For this section and the next section, we will normalize by the minor radius to be consistent. It is clear from these figures that equilibria with improved \minlgradbovera{} have both improved \mindcsovera{} and improved \mindccovera{} at low enough normal field error. This shows the value in using \minlgradb{} as a term in the objective function - the optimized coils have favorable engineering qualities. It is worth noting that for the majority of the rounds of optimization, the coil-coil distance threshold penalty remains inactive. It is only active during the final rounds of optimization, when the coils are long enough that the coils get quite close together. This can be seen in figure \ref{fig:QH_stage2_cc} in the absence of coil sets much below \mindccovera{} = 0.4. 
\begin{figure}[tbp]
     \centering
     \includegraphics[width= \linewidth]{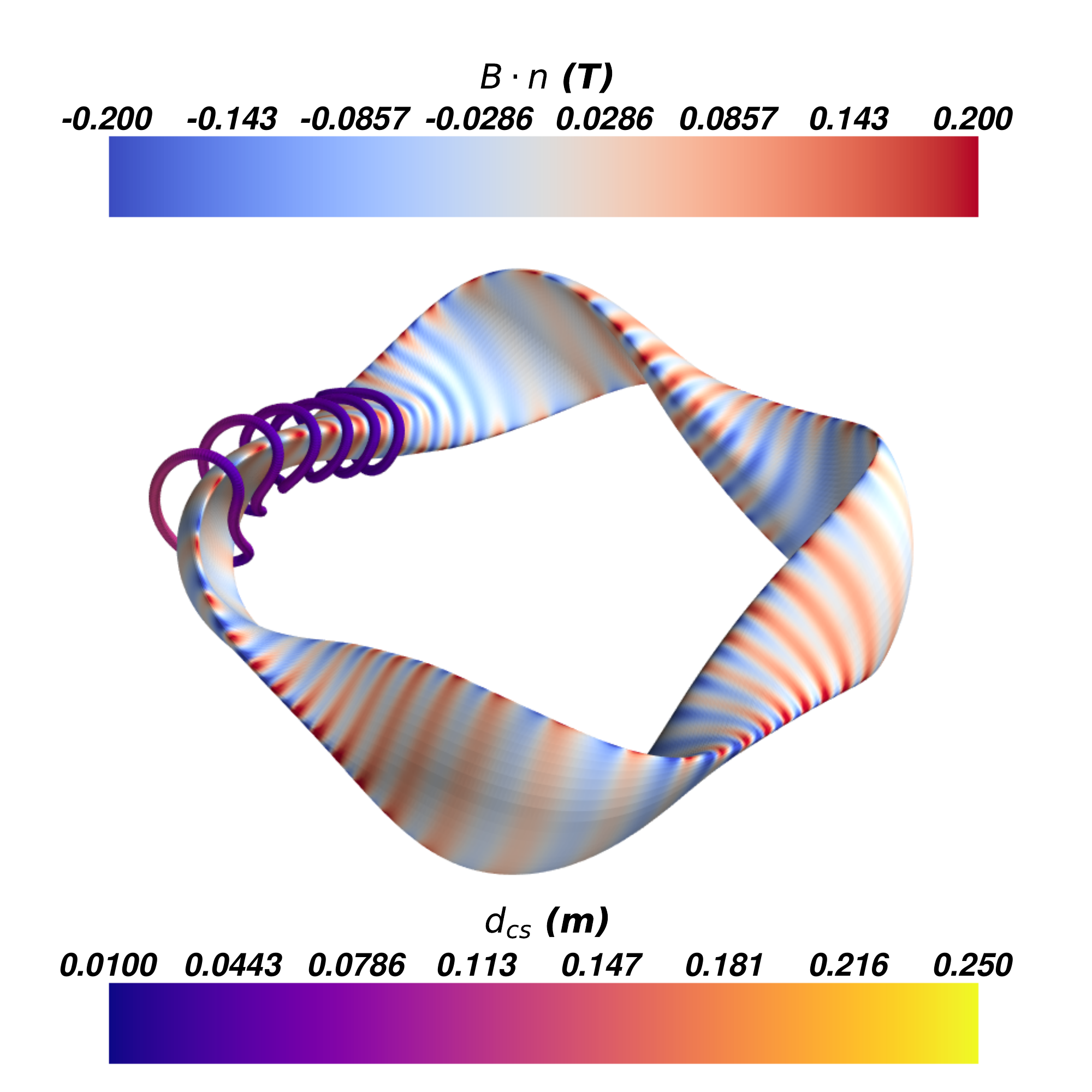}
     \hfill
     \caption{The 6 unique coils optimized for a QH plasma with \minlgradbovera{} =2.31. The length per half field period is 8.602 meters, and \mindcsovera{} = 0.0214. Only the coils for a half field period are shown. The coils are colored to show the distance from the coil to the closest point on the LCFS. The surface is colored to show normal field error. Because the coils are so short, the normal field error - shown in blue and red - form a poloidally closing, striped pattern (coil-ripple.)}
    \label{fig:short_coils}
\end{figure}

\begin{figure}[tbp]
     \centering
     \includegraphics[width= \linewidth]{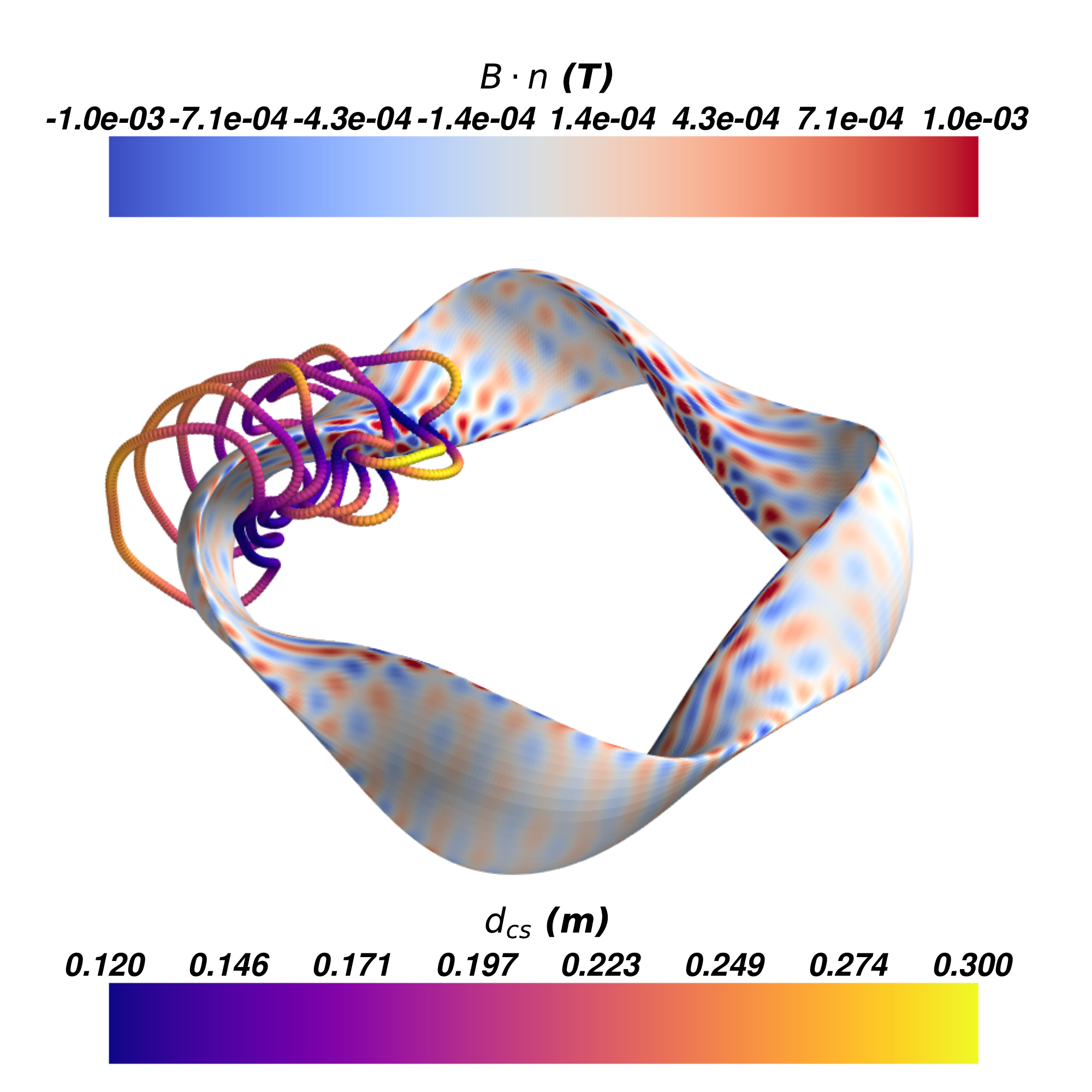}
     \hfill
     \caption{The 6 unique coils optimized for a QH plasma with \minlgradbovera{} =2.31. The length per half field period 19.36 meters and \mindcsovera{}= 0.103. Only the coils for a half field period are shown. The coils are colored to show the distance from the coil to the closest point on the LCFS. The surface is colored to show normal field error. Compared to the previous figure, there is lower normal field error due to coil-ripple (the max \bdotn{} error is over 2 orders of magnitude lower). Because of the constraint on coil plasma distance, the minimum coil-surface distance will only increase until normal field error due to ripple is of similar magnitude to normal field error not due to ripple. At this critical coil length and coil lengths above it, any additional length will be used in order to reduce higher order normal field error. The coils here will become increasingly non-planar. Notice that this happens on the inboard side.}
    \label{fig:long_coils}
\end{figure}
\subsection{Behavior of Coils Close to LFCS and far from LFCS}

It is worth describing these results in more detail for a single equilibrium. The following figures will be for an equilibrium with \minlgradbovera{} of 2.31, but these trends are similar for all equilibria in this dataset. The coils are initialized at the minimum coil length possible that allowed for 6 coils to encircle the plasma and have at least 5 cm between each coil. For this equilibrium, the perimeter of the bean shaped cross section is larger than the perimeter of the triangular cross section. As a result, this means that the coils are too short to encompass the bean shape, and will be bunched up around the section of the plasma with the smallest perimeter. Therefore, the coils are at first optimized with a very small minimum coil-coil distance. These coils and the associated normal field error on the plasma are shown in fig \ref{fig:short_coils}, which corresponds to the orange point in the top left hand corner of fig \ref{fig:QH_stage2_cs}. Because the coils are so close to the plasma, the normal field error is dominated by ripple. In other words, when measuring \bdotn{} on the surface of the plasma, it makes a poloidally striped pattern - in the same direction as the poloidally closing modular coils. Therefore, we will define configurations that have short coils and high normal field error as being in the \textit{high ripple regime}. The normal field error can be reduced by moving the coils farther away. Therefore, whenever the length target is increased, the coils will move farther away from the plasma in order to reduce \bdotn{}. In addition, if there are no coils close to the cross section with the largest perimeter, that cross section will be the location of the largest source of \bdotn{} error. The coils will spread apart from each other in order to reduce \bdotn{} error. These mechanisms explain why, in the first few rounds of increasing coil length, the coils move farther away from the plasma and why the minimum coil-coil length increases. In the high ripple regime, the most important metric that determines \mindcs{} is the difference between length per coil and perimeter around the cross sections nearest to the coil, not \minlgradb{}. In figure \ref{fig:QH_stage2_cs}, since all perimeters of the equilibria are similar, all the equilibria follow a similar "path of optimization" for a \mindcsovera{} between 0 and 0.5. The same is true for \mindcs{} for the shortest coil lengths in figure \ref{fig:QH_stage2_cc}. For equilibria with similar perimeters, increasing \minlgradb{} has a minimal effect on \mindcs{} and \mindcc{} in the high ripple regime.\\
\begin{figure}[tbp]
     \centering
     \includegraphics[width= \linewidth]{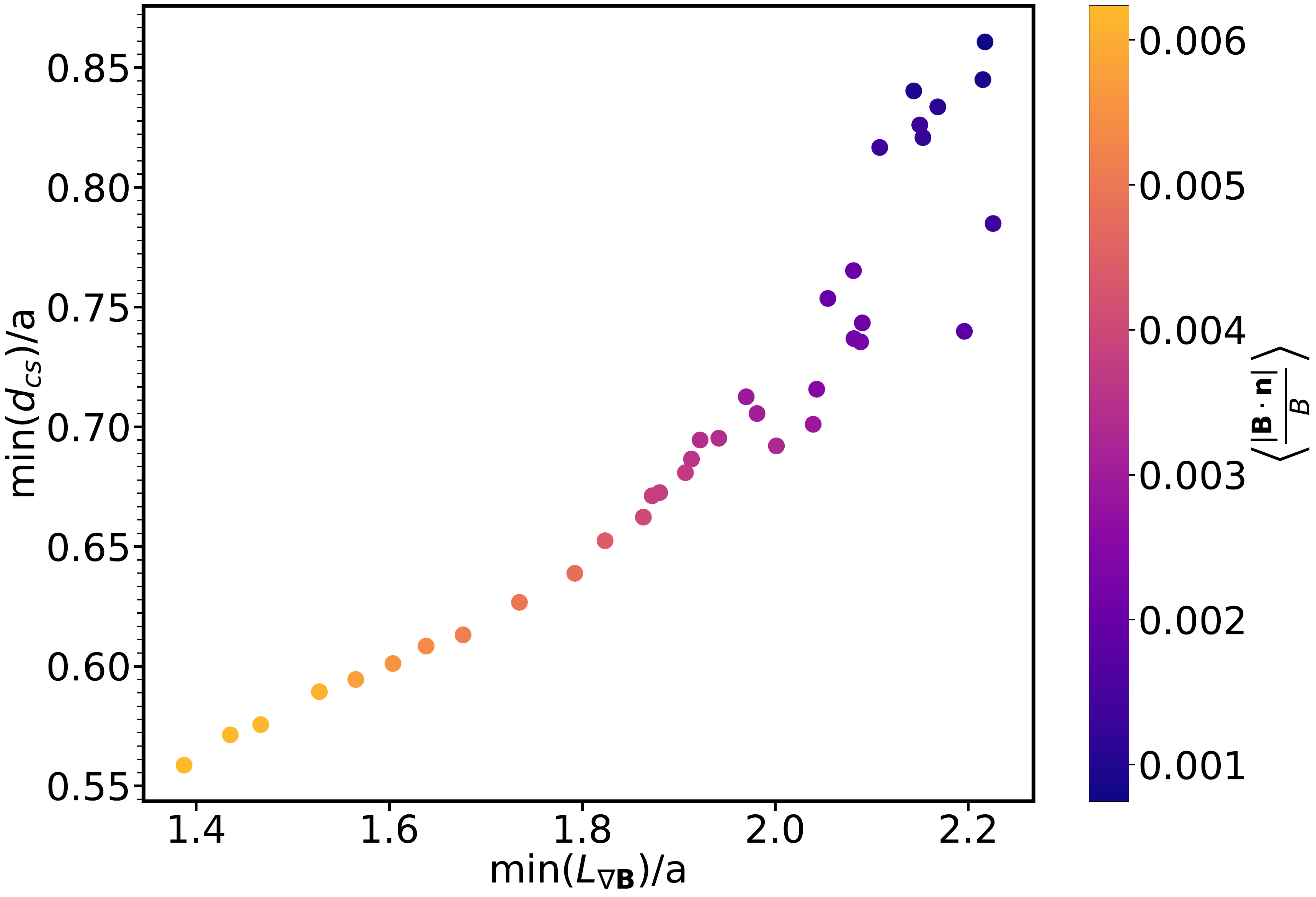}
     \hfill
     \caption{Each dot represents a different coil set for a different equilibrium of the QH family from section \ref{sec:QH}. Each equilibrium has a different \minlgradb{} (shown on the X axis). Each coil set has the same \mindccovera{} $= 0.48$. On the Y axis, the minimum coil-surface distance is shown. In addition, the normalized normal field error is shown on the colorbar. Equilibria with larger \minlgradb{} have a larger minimum coil-surface distance as well as a smaller normal field error.}
    \label{fig:LgradB_vs_cs}
\end{figure}

\begin{figure}[tbp]
     \centering
     \includegraphics[width= \linewidth]{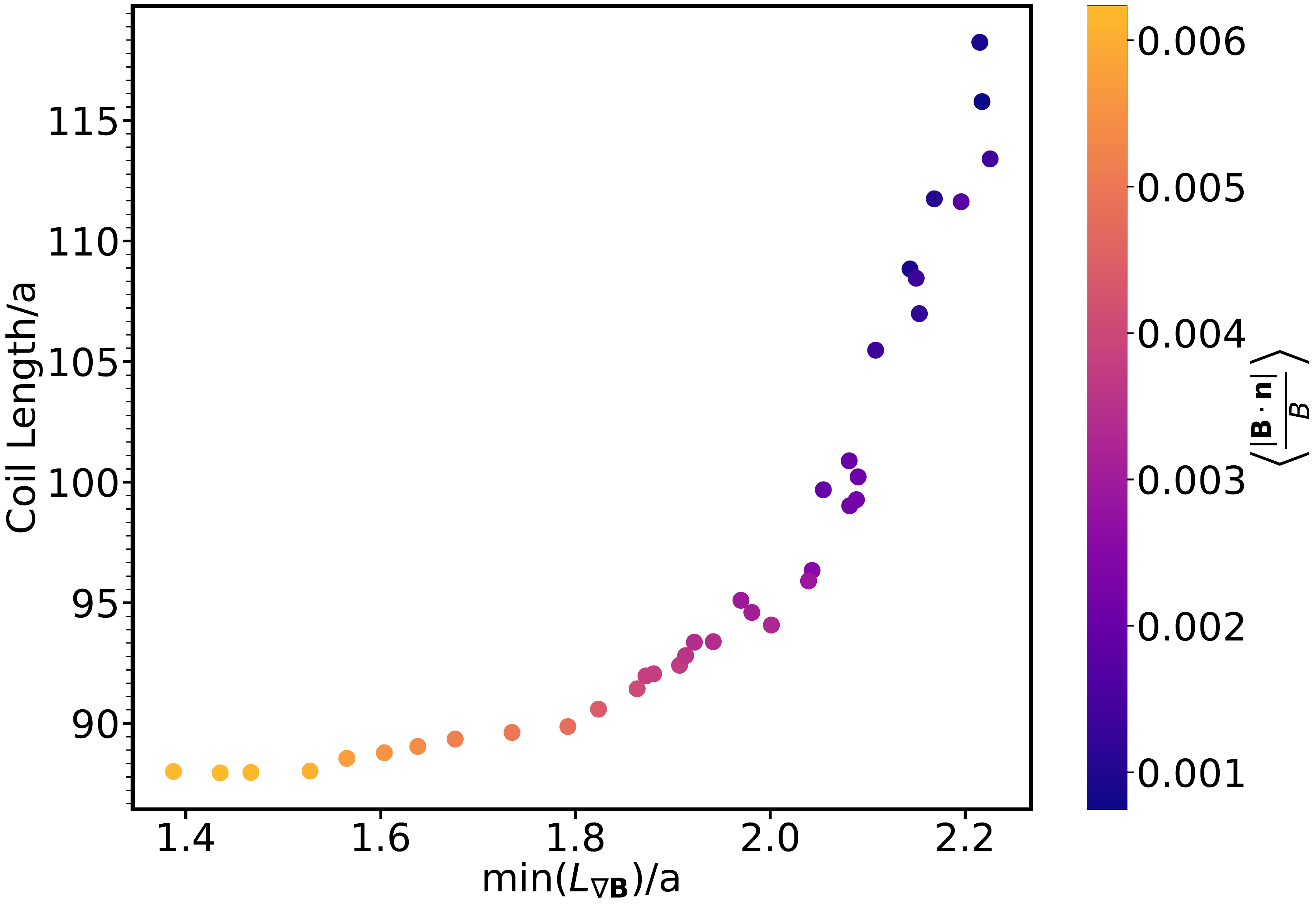}
     \hfill
     \caption{Each dot represents a different coil set for different equilibrium of the QH family from section \ref{sec:QH}. Each equilibrium has a different \minlgradb{} (shown on the X axis). Each coil set has the same \mindccovera{} $=0.48$. This is the same data shown in figure \ref{fig:LgradB_vs_cs}. However, the Y axis now shows the coil length per half field period. Because coil sets with a larger coil-surface distance need greater coil length, equilibria with a larger \minlgradb{} have larger total coil length.}
    \label{fig:LgradB_vs_length}
\end{figure}

Above a critical coil length - which for the configuration in figure \ref{fig:short_coils} is 11.2 meters per half field period - this trend changes. After this point, the \bdotn{} error is no longer ripple dominated everywhere on the plasma. We show the coils and the associated normal field error on the plasma in figure \ref{fig:long_coils}, which corresponds to the yellow star in the bottom right of the figure \ref{fig:QH_stage2_cs}. We call this the low ripple regime. In some locations, the field error will either become helically striped or form a spotted pattern. When this feature occurs, moving the coils farther away from the plasma no longer improves the \bdotn{} error. In the low ripple regime, as coil length is increased, \mindcsovera{} either does not increase (as in the purple curves which have low \minlgradb{}), or it increases more slowly than in the high ripple regime (as in the  orange curves which have high \minlgradb{}). Both scenarios are reflected by the different slope in figure \ref{fig:QH_stage2_cs}. Coil sets that have longer coils - rather than moving farther from the surface - will become more non-planar and decrease \mindcc{}. This is reflected in the curves turning around in the bottom left hand side of figure \ref{fig:QH_stage2_cc}.

\begin{figure}[tbp]
     \centering
     \includegraphics[width= \linewidth]{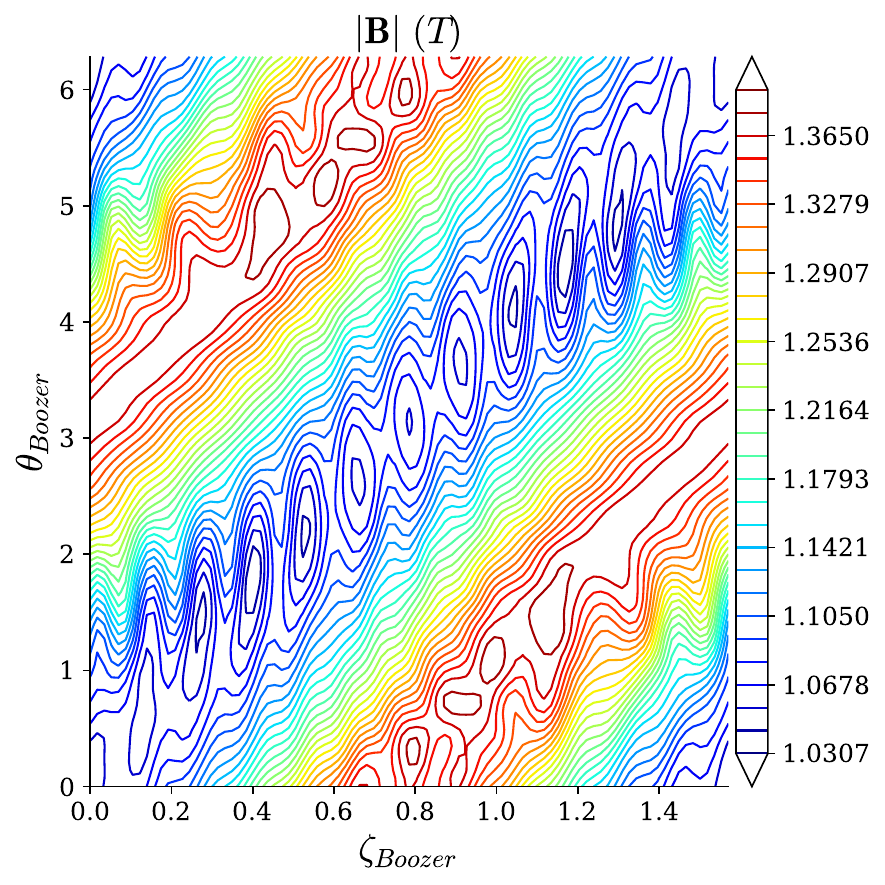}
     \hfill
     \caption{The magnitude of the magnetic field on the free boundary of a QH stellarator with \minlgradbovera{} = 1.35. \mindcsovera{} = 0.069, \mindccovera{} = 0.48 and the length per half field period is 10.88 m. The field is plotted in Boozer coordinates, which means that if the field was quasisymmetric, the contours of constant field strength should be straight lines. This is the same configuration that is shown in figure \ref{fig:stage_1_boozer_low_lgradB}. In this case, the coils that generated the magnetic field are so close to the LCFS that they cause significant coil-ripple, which results in non-straight contours.}
     \label{fig:boozer_lowlgradb}
\end{figure}

\begin{figure}[tbp]
     \centering
     \includegraphics[width= \linewidth]{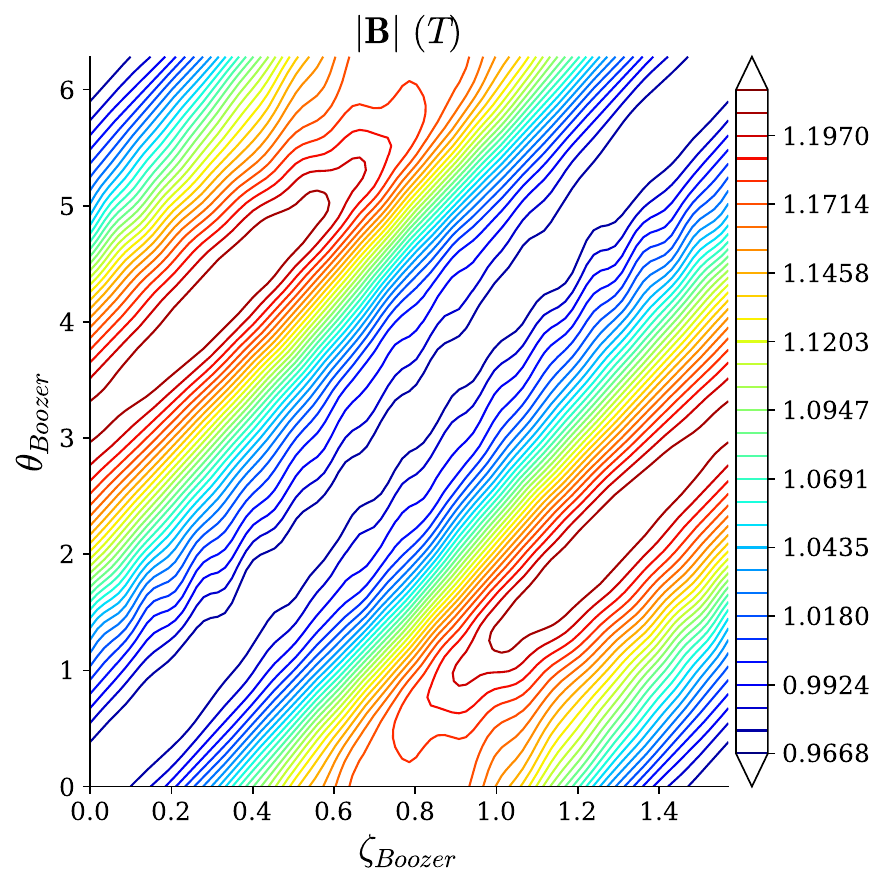}
     \hfill
     \caption{The magnitude of the magnetic field on the free boundary LCFS of a QH stellarator with \minlgradbovera{} = 2.31. \mindcsovera{} = 0.103, \mindcsovera{} = 0.48, and the total coil length per half field period is 14.96 meters. This is the same configuration that is shown in figure \ref{fig:stage_1_boozer_high_lgradB}. For this equilibrium, the coils generating the magnetic field are sufficiently far away that coil-ripple does not affect the field contours as significantly as in figure \ref{fig:boozer_lowlgradb}.}
     \label{fig:boozer_highlgradb}
\end{figure}

For equilibria with similar boundary shapes like this family of QH stellarators, a clear trend can be observed for equilibria with coils of all the same minimum coil-coil distance. The effect is more pronounced at lower minimum coil-coil distances. In the following figures, we filtered for coils with a minimum coil-coil distance between 0.06 m and 0.061 m. We found that for these coil sets, equilibria with a larger \minlgradb{} also had a larger minimum coil-surface distance in a generally monotonic relationship, which is plotted in figure \ref{fig:LgradB_vs_cs}. Of course, since larger coil-surface distances generally need larger lengths, equilibria with larger \minlgradb{} need larger coil lengths as well. This trend is shown in figure \ref{fig:LgradB_vs_length}. The equilibrium with low \minlgradb{} - because the coils are so close to the plasma - is still in the the high ripple regime. To illustrate this, we show a free boundary equilibrium with low \minlgradb{} in figure \ref{fig:boozer_lowlgradb}. This equilibrium has a noticeable ripple of the magnetic field magnitude $B$ on the LCFS when plotted in Boozer coordinates. However, equilibria with high \minlgradb{} are in the low ripple regime. To illustrate this, we show a free boundary equilibrium with high \minlgradb{} in figure \ref{fig:boozer_highlgradb}. It has the same minimum coil-coil distance as the equilibrium shown in \ref{fig:boozer_lowlgradb}, but has much less ripple on $B$ on the LCFS when plotted in Boozer coordinates. Since coil-ripple spoils quasisymmetry, then surprisingly free boundary equilibria with improved \minlgradb{} can have better quasisymmetry, even though the stage-I versions of these equilibria had worse quasisymmetry. This is discussed further in Appendix \ref{app:QS}. This trend impacts the particle confinement as well, shown in the next section.

\subsection{Alpha Particle Confinement Improves with \texorpdfstring{\minlgradb{}}{min LgradB}}

\begin{figure}[tbp]
     \centering
     \includegraphics[width= \linewidth]{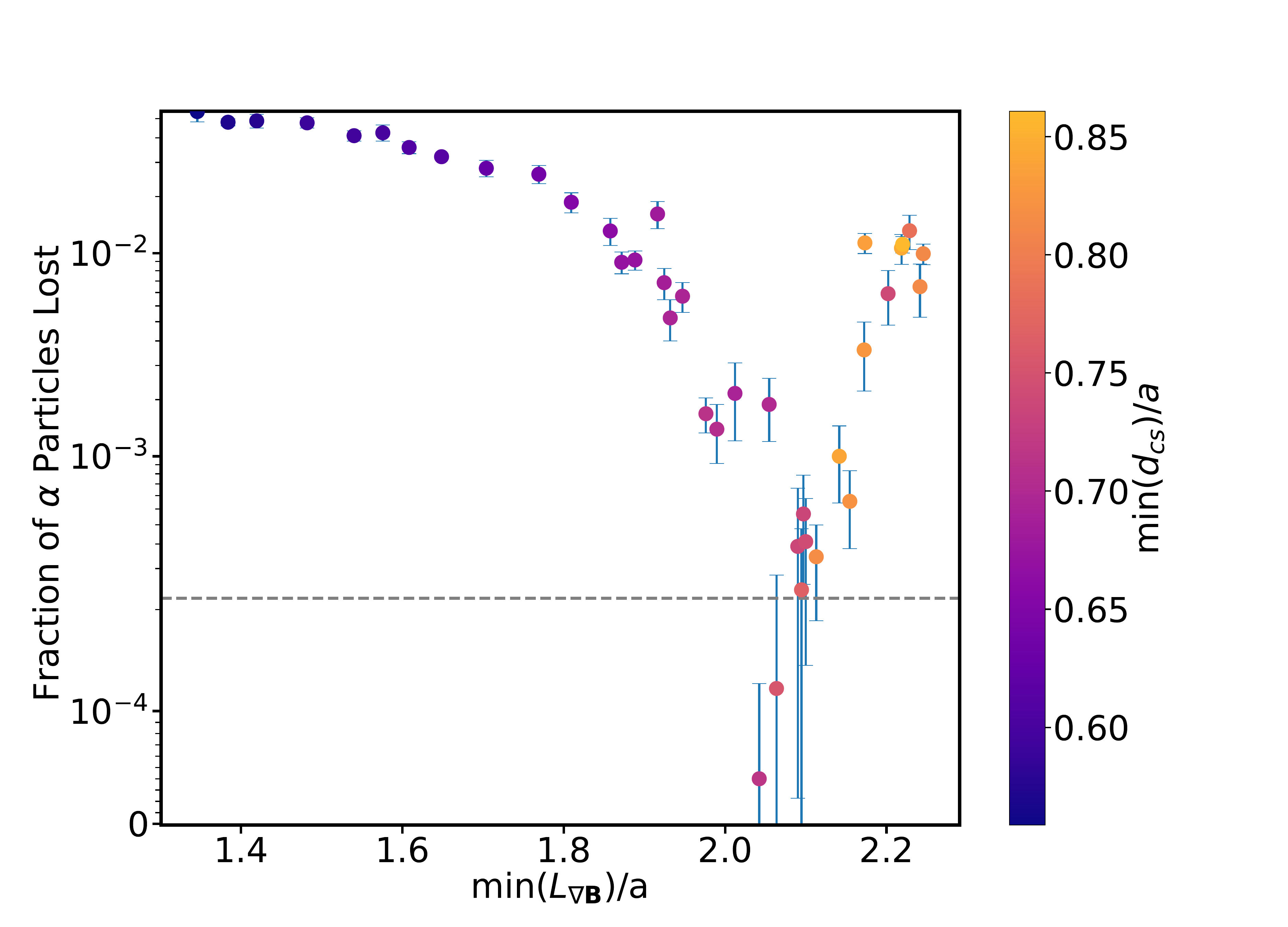}
     \hfill
     \caption{Each point represents a different equilibrium with a different optimized coil set as described in section \ref{sec:QH_coils}. All coils have the same minimum coil-coil distance of 0.05 meters. The code SIMPLE is used to simulate 5000 $\alpha$ particles initialized at s=0.25. The particle loss is shown on the Y axis. 10 runs were simulated, and the standard deviation is shown as error bars. The gray dotted line is located at 1 particle lost per 5000, and data below the line corresponds to average loss of less than one particle per simulation.}
    \label{fig:SIMPLE_alpha}
\end{figure}

We use the guiding center tracing code SIMPLE to trace $\alpha$ particles and measure confinement for the free boundary plasmas.\cite{albert_accelerated_2020}  Each equilibrium was scaled to a minor radius of 1.70 meters and a volume averaged $B$ field of 5.865 T, which are the same as ARIES-CS\cite{najmabadi_aries-cs_2008}. In each configuration, we initialized 5000 particles at  $s = 0.25$ and followed their guiding-center motion for $10^{-1}$ seconds. This is a standard test case used in other papers\cite{landremanMagneticFieldsPrecise2022}. For each geometry, we ran SIMPLE 10 times with different random initial positions and pitch angles (corresponding to isotropic birth), giving a measure of the uncertainty. The results are shown in figure \ref{fig:SIMPLE_alpha}. Equilibria with low \minlgradb{} have worse confinement compared to equilibria with moderately higher \minlgradb{}. This is despite the fact that the quasisymmetry after stage I optimization was better for the low \minlgradb{} equilibria. Configurations with low \minlgradb{} have high coil-ripple which causes significant $\alpha$ loss. For this equilibria family there is a sweet spot where on average less than 1 of the 5000 particles simulated were lost. This occurs around \minlgradbovera{} = 2.04. To understand why this sweet spot exists, refer to the quasisymmetry error shown in figure \ref{fig:stage_1}. Equilibria with high \minlgradbovera{} have poor quasisymmetry, meaning that the fixed boundary equilibrium will have some $\alpha$-particle loss. However, equilibria with low \minlgradb{} have high coil-ripple, which spoils their quasisymmetry following stage 2 (as shown in figure \ref{fig:all_QS_error} in appendix \ref{app:QS}). Configurations with moderate improvement in \minlgradbovera{} have lower quasisymmetry error following stage I than configurations with very large \minlgradb{}, and have better quasisymmetry error following stage II due to lower normal field error. This balance leads to better $\alpha$ particle confinement in the sweet spot.

\section{Finite \texorpdfstring{\protect$\beta$}{beta} Equilibria with Random Boundary Shapes} \label{sec:RAND}

\begin{figure}[tbp]
     \centering
     \includegraphics[width= \linewidth]{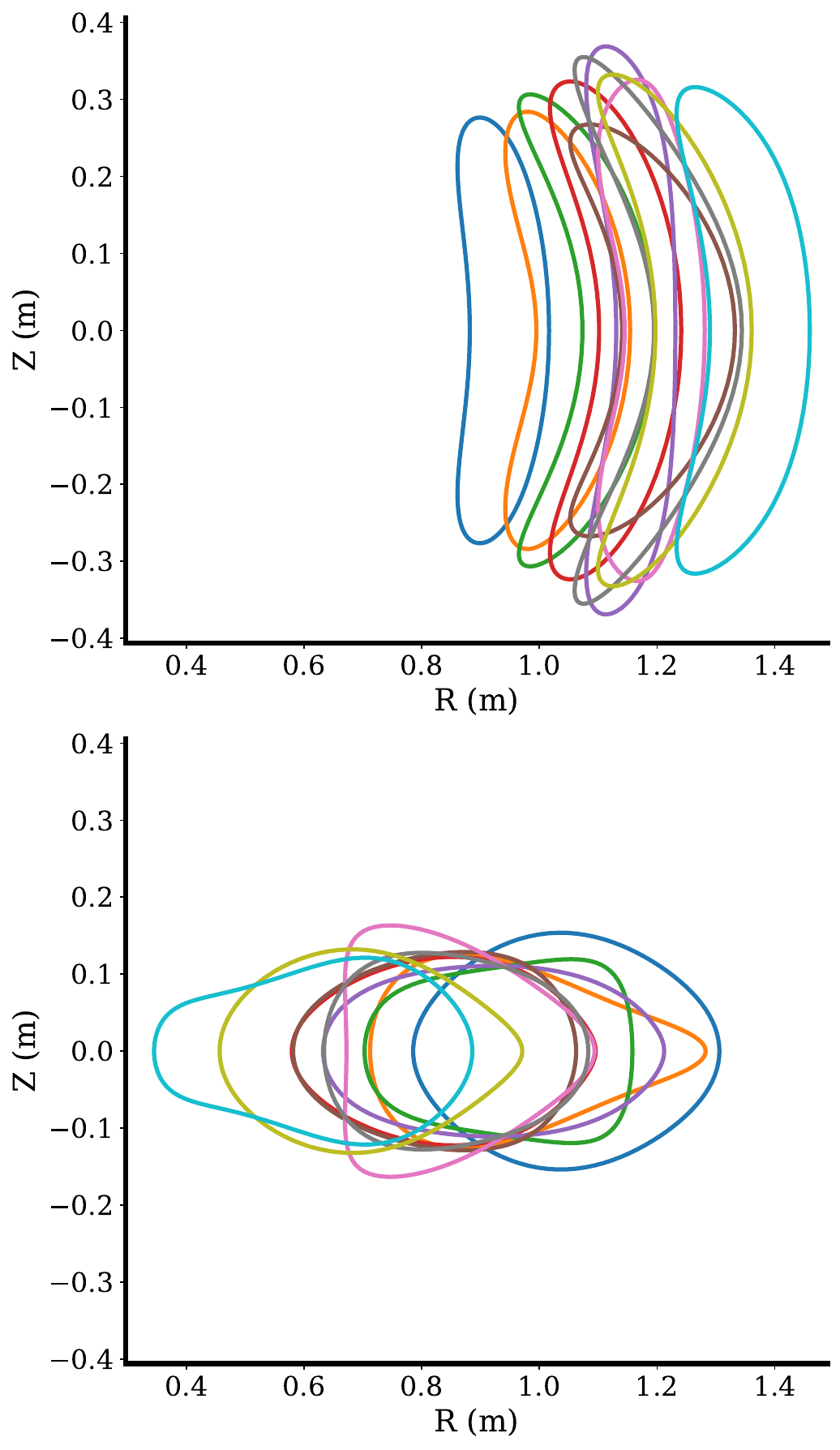}
     \hfill
     \caption{Cross sections of constant $\phi$ for 10 of the 98 configurations from section \ref{sec:RAND}. Shown on the top is the $\phi = 0$ cross section (often referred to as the bean cross section). Shown on the bottom is the $\phi = \pi/2$ cross section (often referred to as the triangular cross section). Compared to figure \ref{fig:bean_and_triangle}, the boundary shapes are more varied.}
    \label{fig:rand_bean_and_triangle}
\end{figure}
\begin{figure}[tbp]
     \centering
     \includegraphics[width= \linewidth]{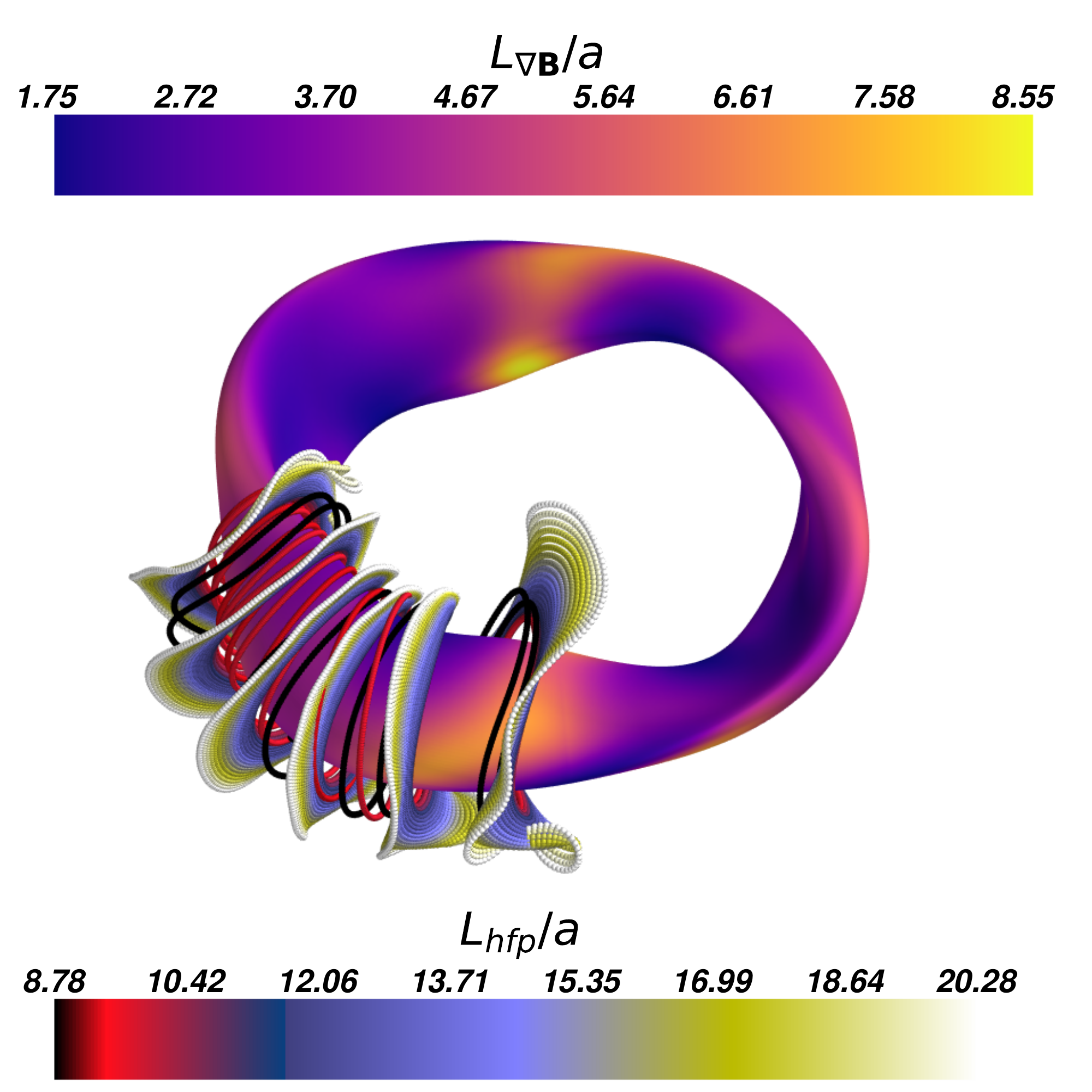}
     \hfill
     \caption{A plasma equilibrium and a sequence of coils optimized using the length continuation method. The configuration is one of the random boundary shapes from the dataset in section \ref{sec:RAND}. The equilibrium has a \minlgradbovera{} of 1.75. The surface is colored with \minlgradb{} at each point. For the coils, each color represents the length of all the coils per half field period normalized by minor radius ($L_{hfp}/a$). Each subsequent coil set uses the previous solution as a warm start, and is optimized with a larger maximum total length. After each round, coils move further from the plasma, but stay close to the plasma at places where \minlgradb{} is small, on the inboard side.
     }
    \label{fig:rainbow_coils} 
\end{figure}

In this section, we look at a dataset of equilibria and associated coils that have different properties than the ones in sections \ref{sec:QUASR} and \ref{sec:QH}. First, the equilibria in sections \ref{sec:QUASR} and \ref{sec:QH} are vacuum fields, while here we will consider finite-$\beta$ configurations. Second, the dataset from section \ref{sec:QUASR} has a large variation in aspect ratio, which influences \mindccovera{}, \mindcsovera{}, and \minlgradb{}; in the present section we will consider configurations that all have the same aspect ratio.
Third, the configurations in section \ref{sec:QH} are all similar to each other, meaning the boundary shapes do not vary significantly. While we had shown that \lgradb{} does correlate with \mindcs{} when boundary shapes are similar, it would be beneficial to see if the correlation remains when the boundary varies significantly. For this dataset, we chose a subset of equilibria previously discussed in \onlinecite{landreman_how_2025,landreman_data_2025}. These equilibria have randomly chosen boundary shapes. The boundaries are represented using a double Fourier series:
\begin{gather}
    R(\theta,\phi) = R_{m,n} \sum_{m,n}\cos(m\theta - n_{fp} n \phi) \\
    Z(\theta,\phi) =Z_{m,n} \sum_{m,n}\sin(m\theta - n_{fp} n \phi),
\end{gather}
where $\phi$ and $\theta$ are the toroidal and poloidial coordinates. The amplitudes $R_{m,n}$ and $Z_{m,n}$ are chosen randomly from a series of normal distributions with the mean and standard deviation matched to a dataset of 44 stellarators.\cite{kappel_magnetic_2023} This dataset gave equilibria of finite $\beta$, an average major radius $R_0$ of 1 meter, and $n_{fp}$ between 2 and 5. Because these boundary shapes are random, the fields are not quasisymmetric. Starting from the original dataset, we filtered for $n_{fp}=2$, an aspect ratio of 6.0, and an average $\iota$ between 0.445 and 0.455. These filters were chosen so that we could maximize the range of \minlgradbovera{} within our dataset. We noticed a trend that there was a larger variation of \minlgradb{} with smaller numbers of field periods, while \minlgradb{} was less varied (and on average lower) with a higher number of field periods.  We chose this range of average iota because it encompassed the largest variation in the minimum \lgradb{}. After filtering, we were left with 98 configurations with a minimum \minlgradbovera{} between 0.56 and 2.95, and volume averaged $\beta$ between 0.02\% and 3.46\%. In figure \ref{fig:rand_bean_and_triangle}, we have plotted the bean and triangle cross sections for 10 of these 98 configurations in order to illustrate the variety in boundary shapes compared to the boundary shapes in figure \ref{fig:bean_and_triangle}. For each equilibrium, coils were optimized at a range of lengths, and 6 coils were optimized per half field period.  During coil optimization, we used a similar objective function to the one used in section \ref{sec:QH_coils}. The only changes were that equation \ref{eq:l} has been replaced by $f_{l,i} =  \max\bigl(0, \bigl(\sum_{n} l_n-l_i^*\bigr)^2 \bigr)$, and that $\mathbf{B_{plasma}}$ is not zero. We used the virtual casing method in \texttt{SIMSOPT} to calculate it. This means that - for this section - we also calculate \lgradb{} from $B_{coils}$ and $\mathbf{\nabla B_{coils}}$ (the magnetic field from the coils only, not the total magnetic field). In figure \ref{fig:rainbow_coils}, we show the optimized coils for various length targets for one equilibrium.

\begin{figure}[tbp]
     \centering
     \includegraphics[width= \linewidth]{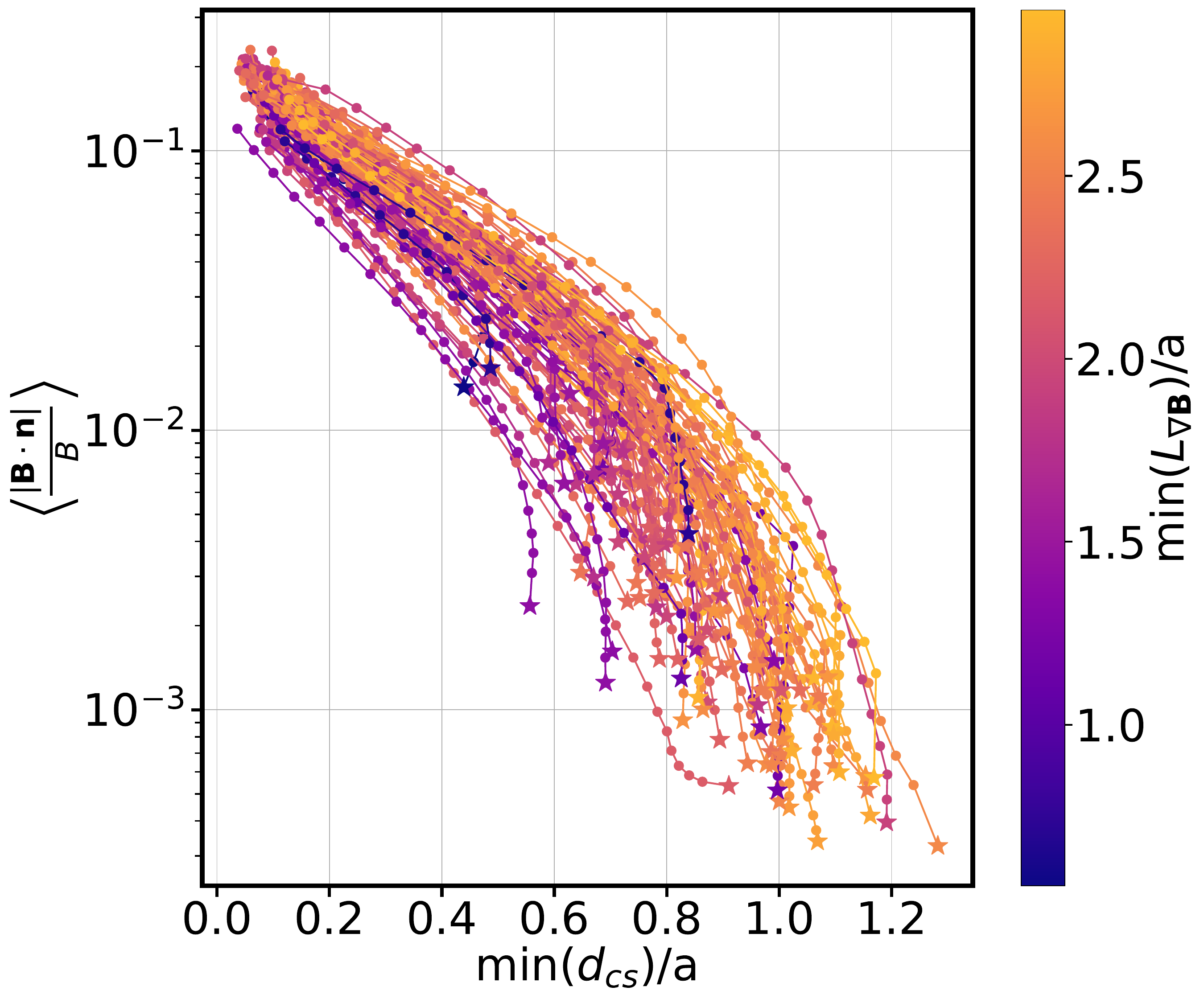}
     \hfill
     \caption{The data related to a series of stage II coil optimizations are plotted. Each colored curve represents a different equilibrium with a randomized boundary shape and with a different minimum \minlgradbovera{}. These configurations are the ones discussed in section \ref{sec:RAND}. Starting in the top left corner and moving down right, each dot represents a coil set optimized by a subsequent round of the continuation method, where the objective function is modified to increase coil length. The previous solution is used as a warm start for the next dot. Each star represents the final coil set with maximum coil length. There is a trend that equilibria with better minimum \minlgradbovera{} have improved minimum coil-surface distance (on the X axis) and absolute normal field error (on the Y axis).}
    \label{fig:rand_scan_cs}
\end{figure}

\begin{figure}[tbp]
     \centering
     \includegraphics[width= \linewidth]{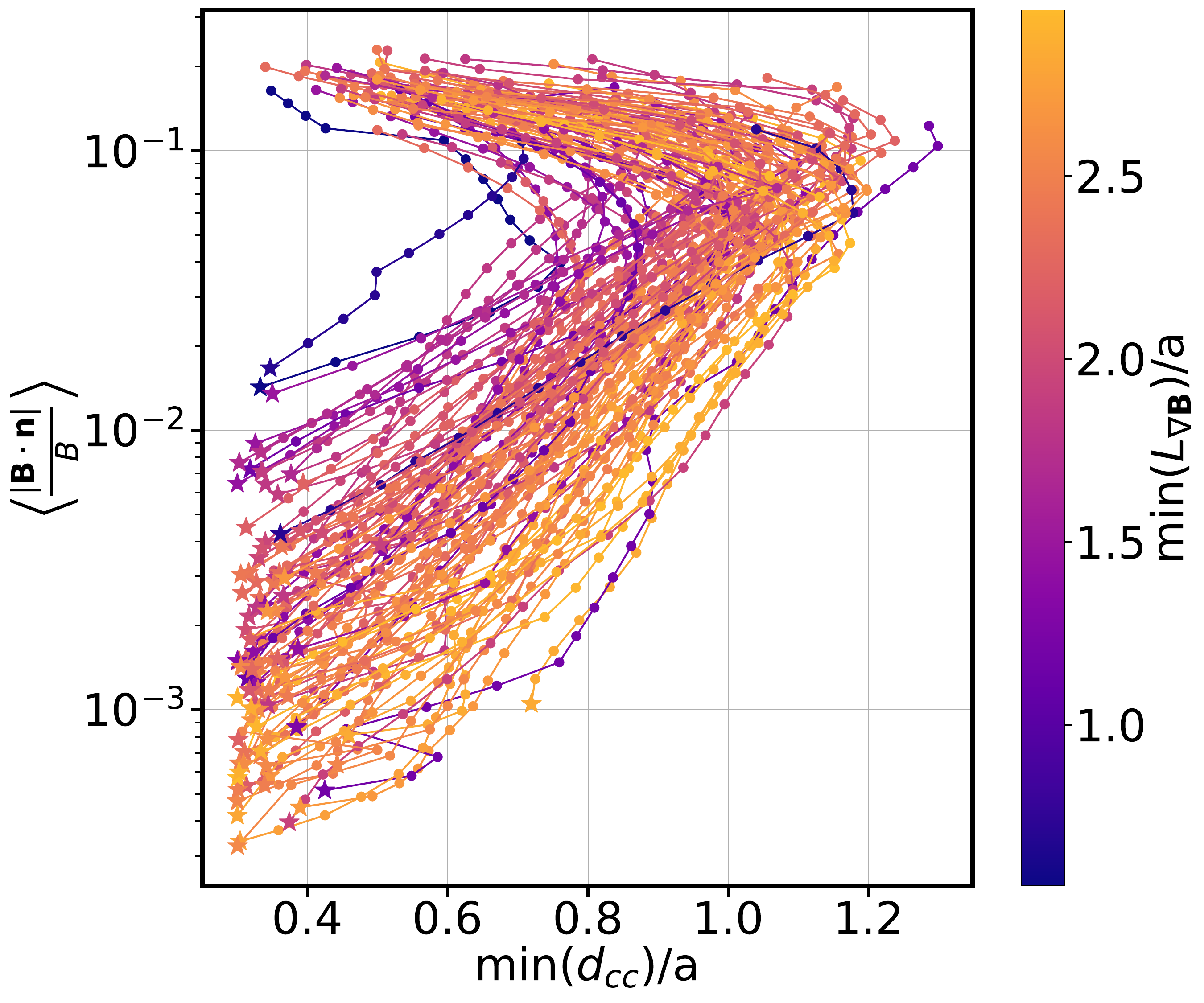}
     \hfill
     \caption{Sequence of stage 2 coil optimization using a continuation method, from section \ref{sec:RAND}. This is the same data shown in figure \ref{fig:rand_scan_cs}. Each colored curve represents a different plasma equilibrium from the randomized equilibria set with a different minimum \lgradb{}. Starting in the top left corner and moving downward, each dot represents a coil set optimized by a subsequent round of the continuation method. The objective function is modified and the previous solution is used as a warm start for the next dot. This method ensures improvements in normal field error and coil-surface distance as the length afforded to the coils is increased. In addition, there is a trend that equilibria with better minimum \lgradb{} have improved minimum plasma coil distance and normal field error.}
    \label{fig:rand_scan_cc}
\end{figure}

\begin{figure}[tbp]
     \centering
     \includegraphics[width= \linewidth]{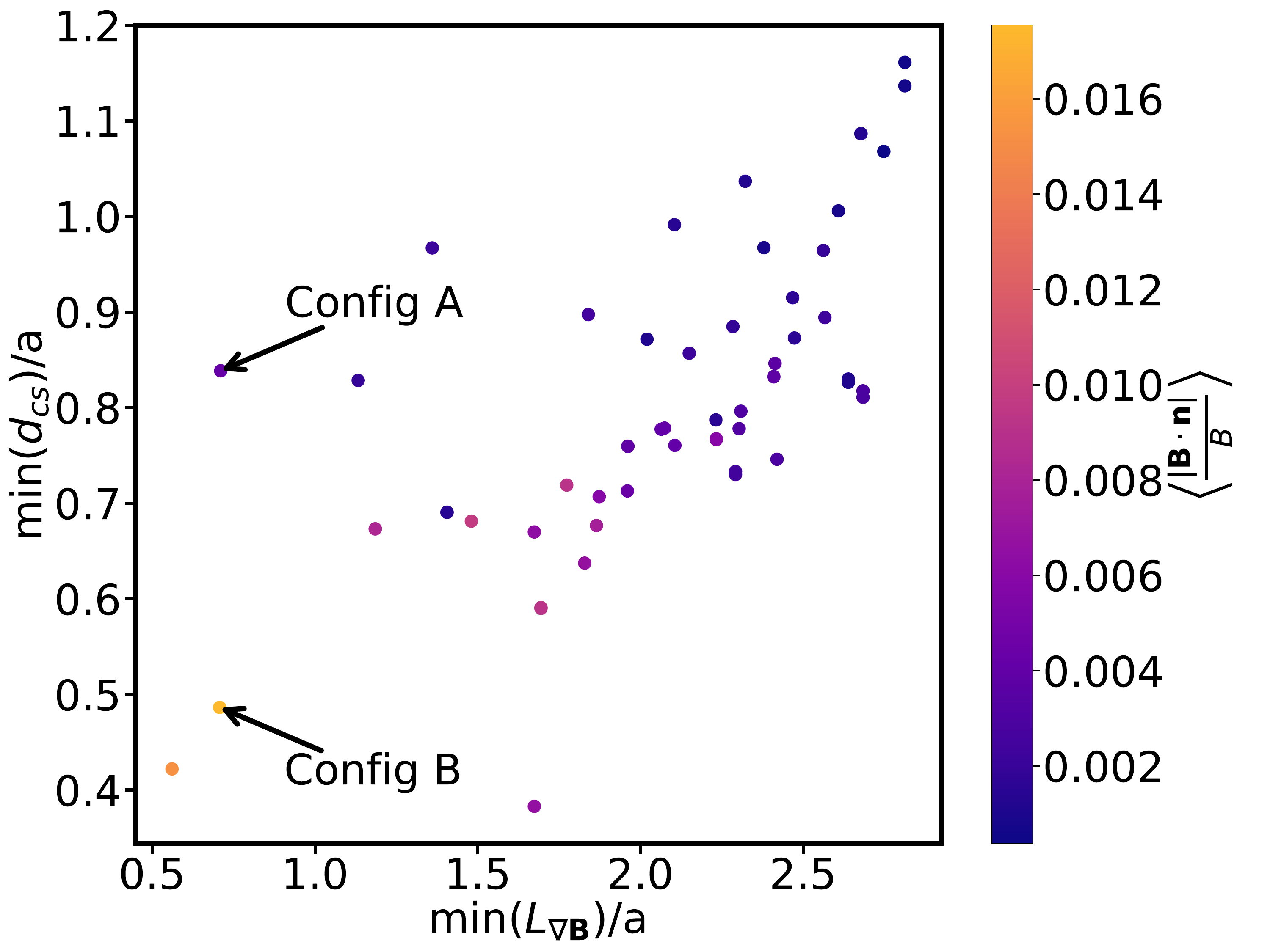}
     \hfill
     \caption{Each dot represents a different coil set for a different equilibrium with a randomized boundary shape from section \ref{sec:RAND}. Each configuration has a different \minlgradb{} (shown on the X axis). Each coil set has the same \mindccovera{} at 0.36. On the Y axis, the minimum coil-surface distance is shown. There is a positive correlation between \minlgradbovera{}{} and \mindcsovera{}. Configurations A and B have nearly the same \minlgradbovera{} (0.710 and 0.707 respectively) and nearly the same \mindccovera{} (0.361 and 0.360 respectively), but a large difference in \mindcsovera{} (0.839 and 0.486 respectively).}
    \label{fig:rand_lgradb_vs_cs}
\end{figure}

The results of the optimization are shown in figures \ref{fig:rand_scan_cs}, \ref{fig:rand_scan_cc}, and \ref{fig:rand_lgradb_vs_cs}. These figures are analogous to figures \ref{fig:QH_stage2_cs}, \ref{fig:QH_stage2_cc}, and \ref{fig:LgradB_vs_cs}, respectively.  Each figure for the random-boundary configurations has a similar trend to the corresponding figure from section \ref{sec:QH}, but with more scatter.  As in section \ref{sec:QH}, increased \minlgradb{} is correlated with improved \mindcs{}, \mindcc{}, and \bdotn{}. The weaker correlation compared to the figures in section \ref{sec:QH} is mostly likely due to the diversity of boundary shapes in this dataset.

Another factor that may explain the greater scatter in the data for this section compared to section \ref{sec:QH} is the greater variation in the location on the LCFS of \lgradb{}, as this location may also impact \mindcs{}. Consider two equilibria with nearly the same \minlgradb{}, labeled A and B on figure \ref{fig:LgradB_vs_cs}. The also have almost the same \mindccovera{}, but the \mindcsovera{} values are quite different, 0.839 and 0.486 respectively. Equilibrium A has \minlgradb{} on the outboard side, while Equilibrium B has \minlgradb{} on the inboard side. Considering only \lgradb{}, we might expect both equilibria to have the same minimum coil-surface distance. However, the coil-coil distance gets larger on the outboard side in equilibrium A as \mindcs{} increases, while coil-coil distance gets smaller on the inboard side on equilibrium B with increasing \mindcs{}. There is a maximum coil-surface distance that the coils can achieve before reaching this coil threshold $d_{cc}*$. On the outboard side, the limiting factor on coil-surface distance is \minlgradb{}, while the limiting factor on the inboard side can either be \minlgradb or $d_{cs}*$. Further research could be done into this hypothesis, and into how to more quantitatively relate the location and value of \minlgradb{} and to \mindcs{}.

Overall, this dataset serves to support the robustness of \minlgradb{} as a proxy for \mindcs{}. These quantities still show moderate correlation even for a diverse set of plasma boundary shapes and even when $\beta$ is finite. This analysis was repeated using both four and three coils per half field period (not shown), and similar trends were found.

\section{Conclusion}

In this paper, we have presented extensive evidence that \minlgradb{} is predictive of \mindcs{} for filamentary coils, building on our previous analysis with the current potential model \cite{kappel_magnetic_2024}.
The analysis here included three datasets with different properties. First, we examined existing coils and equilibria from the QUASR dataset generated via single stage optimization. We found good correlation between \mindcs{}  and \minlgradb{}, and even better between \mindcs{} and \minlgradgradb{}. We also found that for any one configuration, the locations of \minlgradbovera{} and \minlgradgradbovera{} on the plasma boundary usually are close to the location of \mindcs{}. Second, we optimized a family of QH stellarators for improved \minlgradb{}, and optimized multiple coil sets for each with varying coil lengths. We found a general trend that equilibria with larger \minlgradbovera{} had better \mindcsovera{}, and that trend was more clear when \mindcc{} was held fixed. Surprisingly, \minlgradb{} is \textit{not} correlated with \mindcs{} when the coil length is so low that the resulting configuration is in the high ripple regime, as shown in figure \ref{fig:short_coils}. Because larger \minlgradb{} results in reduced coil-ripple at the same \mindcc{}, optimizing for higher \minlgradb{} can reduce the loss of $\alpha$ particles, up to point. There exists a ``happy medium'' where both the quasisymmetry error in stage I is low (at low \minlgradb{}) and the ripple error is low (at high \minlgradb{}), as shown in figure \ref{fig:SIMPLE_alpha}. In other words, optimizing \minlgradb{} up to a point in stage I does result in improved coils and confinement following stage II. Finally, we optimized multiple coil sets for equilibria with random boundary shapes in order to see if \minlgradb{} alone is sufficient to determine \mindcs{}. While \minlgradb{} did have moderate correlation with \mindcs{}, additional research should be done on other metrics related to the magnetic field structure and boundary shapes. It is worth noting that while the QUASR and random-boundary datasets had moderate rather than perfect correlation between \minlgradb{} and \mindcs{}, this only suggests that \minlgradb{} has limited correlation when comparing configurations with very different shapes. When shapes are similar, such for the QH dataset, \minlgradb{} is an effective proxy for \minlgradb{}. We believe this means it is more effective as an objective function for optimization than as a comparison metric between widely different configurations.

This study invites further research into the factors that determine \mindcs{}. While \lgradb{} acts as a good first order approximation, additional research could be done into corrections that would improve the correlation, such as the perimeter of constant-$\phi$ curves of the LCFS. At least for the QUASR dataset, \mindcs{}  shows surprisingly strong correlation with \minlgradgradb{}, and additional research should be done to see how well \minlgradgradb{} works as an optimization metric.
It may turn out that it is a poor objective as computing it accurately requires high resolution which may slow optimization. One reason \minlgradgradb{} may be better correlated with \mindcs{} than \minlgradb{} is that \minlgradb{} provides very local information, while \minlgradgradb{} provides more information about how $\mathbf{B}$ varies away from the evaluation point. Using other reductions than the minimum may also provide some more global information, such as using quantiles or the harmonic average of \minlgradb{}. In addition, it would be useful to add mathematical rigor to quantify \lgradb{}'s effect on coil complexity. For example, research by Golab \textit{et al.}\cite{golabTwoDimensionalHarmonicExtensions2023,golabStellaratorCoilsExtensions2023} approaches a similar problem by determining how far a vector field can be harmonically extended from the boundary of a domain. This determines how far a current-carrying coil could be from the LCFS. While those studies explored both two and three dimensions, additional research would be needed to improve the tightness of the bounds for practicality.

\section*{Data Availability}

Data associated with this study can be downloaded from Zenodo : \href{https://zenodo.org/records/18726860}{10.5281/zenodo.18726860}.\cite{kappelSupplementalFilesHow2026}

\section*{Acknowledgments}

We would like to thank Andrew Giuliani, Rory Conlin, and Misha Padidar for their thoughtful discussions that helped our research.

\section*{Funding}

This work was supported by the U.S. Department of Energy, Office of Science, Office of Fusion Energy Science, under award number DE-FG02-93ER54197. This research used resources of the National Energy Research Scientific Computing Center
(NERSC), a U.S. Department of Energy Office of Science User Facility located at Lawrence Berkeley National Laboratory, operated under Contract No. DE-AC02-05CH11231 using NERSC award FES-ERCAP-mp217-2025.

\section*{Funding}

M. L. is a consultant for Type One Energy Group.

\appendix
\section{\texorpdfstring{Location of the Largest Norm of the $\nabla \mathbf{B}$ Tensor}{Location of the Largest Norm of the grad B Tensor}}
\label{app:gradB}

In the body of this paper, we stated that the largest value of $\| \nabla \mathbf{B}\|_F$ for a vacuum field is achieved on the LCFS. Here we prove this fact. 
The reasoning is similar to the proof in Ref. \onlinecite{rodriguez2024maximum} that $B$ is maximized on the LCFS.
In vacuum, $\nabla \mathbf{B} = \nabla \nabla \Phi$, where $\Phi$ is a scalar magnetic potential that satisfies $\nabla^2 \Phi = 0$. We next prove that $ \| \nabla \mathbf{B}\|_F^2$ is a subharmonic function. 
A function $f(\mathbf{x})$ is defined to be subharmonic if 
\begin{ceqn}
\begin{equation}
    \nabla^2 f(\mathbf{x}) \geq 0
\end{equation}
for all $\mathbf{x}$.
\end{ceqn}
We can evaluate
\begin{ceqn}
\begin{align*}
    \nabla^2 \left( \| \nabla \mathbf{B}\|_F^2 \right)
    &= \partial_k \partial_k  \bigl((\partial_i \partial_j \Phi )( \partial_i \partial_j \Phi )\bigr)\\
        &= 2  \bigl( ( \partial_i \partial_j \underbrace{\partial_k \partial_k \Phi}_{=0} )( \partial_i \partial_j \Phi ) + ( \partial_i \partial_j \partial_k \Phi ) ( \partial_i \partial_j \partial_k\Phi )\bigr)\\
        &= 2 ( \partial_i \partial_j \partial_k \Phi ) ( \partial_i \partial_j \partial_k\Phi ), 
\end{align*}
\end{ceqn}
where the final expression is a sum of squares and hence nonnegative.
Therefore, $\| \nabla \mathbf{B}\|_F^2$ is subharmonic.
The maximum of a non-constant subharmonic function cannot be in the interior of the domain, hence the maximum of $\| \nabla \mathbf{B}\|_F^2$ must be achieved on the LCFS. Since $\sqrt{x}$ is monotonic, $\| \nabla \mathbf{B}\|_F$ must also attain its maximum on the LCFS.

Through the similar proof in Ref. \onlinecite{rodriguez2024maximum}, one can show that $|B|$ is also a subharmonic function and so must have its maximum achieved on the LCFS. We would not expect the ratio between two subharmonic functions to be subharmonic, so it is unclear whether \lgradb{} will always achieve its minimum on the LCFS. 
However the relative variation of $B$ on the boundary is typically much smaller than the variation of $\| \nabla \mathbf{B}\|_F$, in which case it is very likely that \lgradb{} is minimized where $\| \nabla \mathbf{B}\|_F$ is maximized.
Alternatively, if we replace \lgradb{} with $\sqrt{2}\langle B\rangle_{\rho=1}/ \|\nabla\mathbf{B}\|_F$, then the numerator is independent of position, so this scale length will surely achieve its minimum on the LCFS. 

When the plasma $\beta$ is finite, we compute \lgradb{} from only the externally supplied vacuum field using the virtual casing method, so the argument above remains valid.
If \lgradb{} were computed instead for the total field, in our experience the difference from the vacuum \lgradb{} is small (since typically $\beta$ is $\ll 1$), so it is likely that the minimum may still be attained on the boundary.
\begin{figure*}[tpb]
     \centering
     \includegraphics[width= 1 \textwidth]{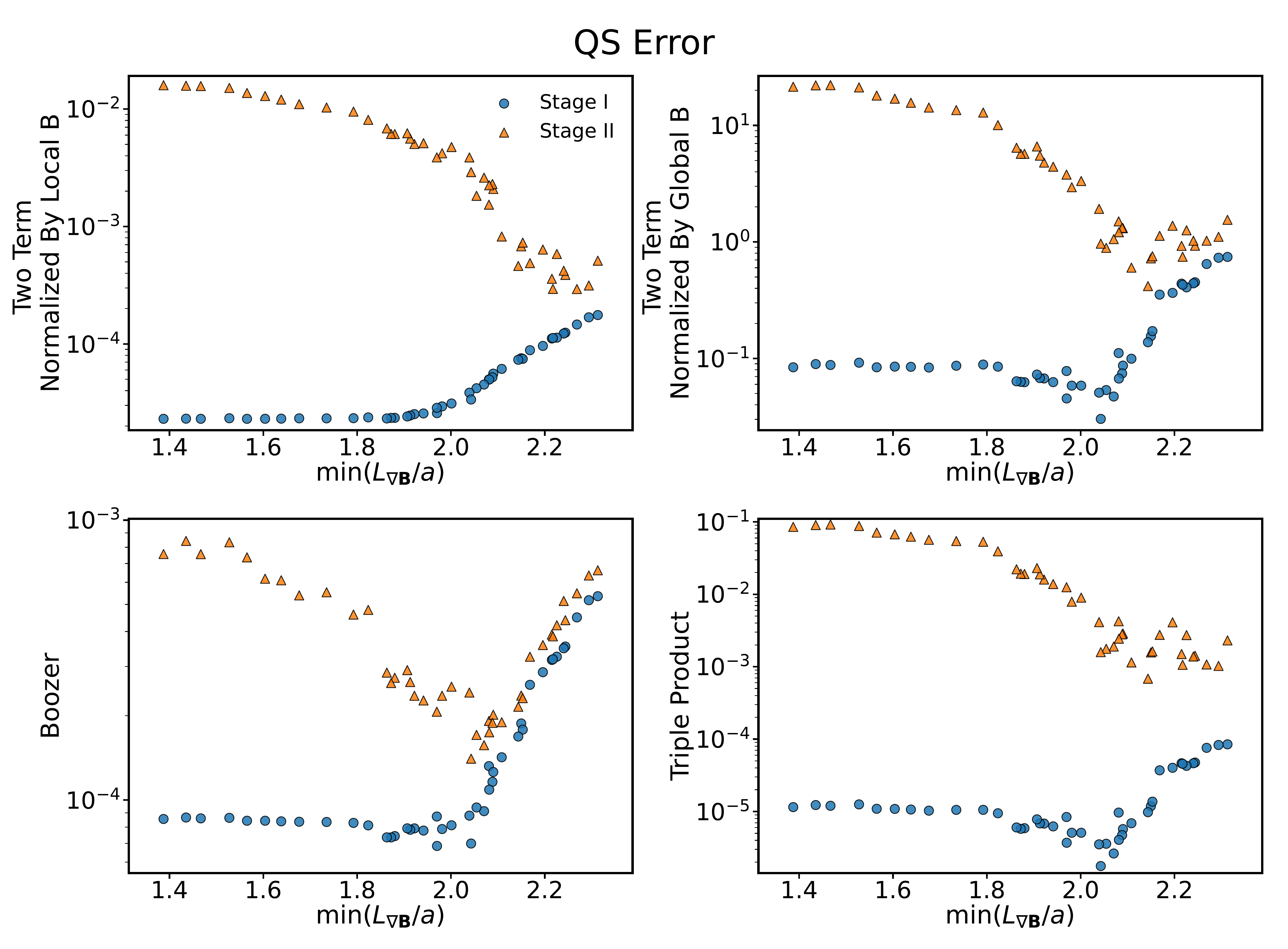}
     \hfill
     \caption{Metrics of quasisymmetry error are shown for the QHH configurations in section \ref{sec:QH}. The blue dots indicate fixed boundary equilibria after stage 1 optimization, and the orange dots represent free boundary equilibria after coils were optimized for them in stage II. The coils all have the same \mindcsovera{}=0.4. While the free boundary configurations will generally have slightly different \minlgradbovera{} before vs. after stage II optimization, we plotted these points using the stage I values for easier comparison. For the Y axes on all plots: the top left depicts the two term quasisymmetry error normalized by the local $B^3$ (which was used in the objective function for the stage I optimizations); the top right depicts the two term quasisymmetry error normalized by a global average $B$; the bottom left depicts the Boozer quasisymmetry error; and the bottom right depicts the triple product error.}
    \label{fig:all_QS_error}
\end{figure*}
\section{How Different Metrics of Quasisymmetry Error Change Between Stage I and Stage II}

\label{app:QS}
In this section, we shall provide some additional figures associated with the quasisymmetry error for the dataset presented in \ref{sec:QH}. 
There are three established measures of quasisymmetry error,\cite{rodriguez2022measures} referred to as Boozer quasisymmetry, two-term quasisymmetry and triple-product quasisymmetry error. To make the two-term expression dimensionless, it is normalized by the cube of a field strength, which could be taken to be the local field strength $B$ (as in the optimizations in section \ref{sec:QH_stageI}) or a global average. While a comparison of the quasisymmetry error for the family of QH equilibria was shown in figure \ref{fig:stage_1}, for completeness’s sake we show all four quasisymmetry error metrics in figure \ref{fig:all_QS_error} (as noted by the blue circles). The key takeaway is that all four metrics generally have the same shape. Above a critical threshold of \minlgradb{}, the quasisymmetry error increases. 

We also show all for metrics for the free boundary equilibria after optimizing coils in stage II. The coil sets used all have a minimum coil-coil distance of 0.05 m, the same set used in figures \ref{fig:LgradB_vs_cs}--\ref{fig:SIMPLE_alpha}. The quasisymmetry errors are also shown in figure \ref{fig:all_QS_error} (as noted by the orange triangles). Three of the quasisymmetry error metrics - two-term with local or global normalization, and triple product - all improve with increasing \minlgradb{}. This is not true for Boozer quasisymmetry error, which dips down and goes back up, similar to the curve of $\alpha$-particle losses in figure \ref{fig:SIMPLE_alpha}. The other metrics place greater weight on quasisymmetry error at high mode numbers \cite{rodriguez2022measures}, such as the coil-ripple that is prominent in the free-boundary low-\minlgradb{} equilibria. This difference explains why the relative increase in quasisymmetry error from stage I to stage II is less for the Boozer quasisymmetry error than for the other three metrics. The different weighting of mode numbers between the four metrics may also be part of the reason that Boozer quasisymmetry error is better correlated with the $\alpha$-particle losses compared to the other metrics.

\nocite{*}
\bibliography{gradientoptimization}

\end{document}